\documentclass{svjour3}                     

\usepackage{lineno,hyperref}
\usepackage{array,multirow,makecell}
\usepackage{url}
\usepackage{xcolor}
\usepackage{algorithm,algpseudocode}
\usepackage{setspace,varwidth}
\usepackage{verbatim}
\usepackage{moreverb}

\usepackage{amsmath,amsfonts,amssymb}
\usepackage{newtxtext}
\usepackage{newtxmath}
\usepackage[mathscr]{eucal}
\usepackage{bm}

\usepackage{morefloats}
\usepackage{stackrel}
\usepackage{caption}
\usepackage{BOONDOX-cal}

\usepackage{gastex}
\usepackage{float}
\usepackage{tikz}
\usepackage{pgfplots}
\pgfplotsset{compat=1.17}
\usetikzlibrary{shapes,decorations,arrows,automata,plotmarks,patterns,petri}
\usepackage{tikz-inet}
\usepackage{pgf}

\usepackage{natbib}

\usepackage{crayola}
\usepackage{mycal}
\usepackage{myboldmath}
\usepackage{mymathbb}

\def\bzeta{{\boldsymbol\zeta}}
\newcommand{\fraku}{{\mathfrak{u}}}

\smartqed  

\def\rqmc{{\rm rqmc}}

\def\eqdef {\buildrel \rm def \over =}    

\def\var{{\rm Var }}
\def\q {$\kern15pt$}    
\def\?{\discretionary{}{}{}}  

\def\Var{{\rm Var}}

\def\tr{{\sf t}}

\def\vrf{\mbox{\sc vrf19}}
\def\eif{\mbox{\sc eif19}}
\def\er{\mbox{\sc eif19}}

\def\camp{\rm cAMP}
\def\pka{\rm PKA}
\def\pkatwo{\rm {PKA\text{-}cAMP}_{2}}
\def\pkafour{\rm {PKA\text{-}cAMP}_{4}}
\def\pkar{\rm PKAr}
\def\pkac{\rm PKAc}

\newif\ifnotes\notestrue
\def\boxnote#1#2{\ifnotes\fbox{\footnote{\ }}\ \footnotetext{ From #1: #2}\fi}

\def\red#1{#1}
\def\new#1{{#1}}

\def\pierre#1{\boxnote{Pierre}{\color{red}#1}}
\def\hpierre#1{}
\newcommand{\hmpierre}[1]{{}}

\def\hamal#1{}

\def\hflorian#1{}

\pgfplotscreateplotcyclelist{defaultcolorlist}{%
	{blue!95!black,line width=0.9pt,mark=*,solid},
	{red!95!black,line width=0.9pt,mark=square,solid},
	{green!98!black,line width=0.9pt,mark=triangle*,solid},
	{black,mark=star,line width=0.9pt,solid},
	{purple!85!black,line width=0.9pt,mark=o,dashed,mark options=solid},
	{magenta!85!black,line width=0.9pt,mark=square,dotted,mark options=solid}
	}
\pgfplotscreateplotcyclelist{varPerStep}{%
	{green!98!black,line width=0.9pt,solid},
	{black,line width=0.9pt,solid},
	{purple!85!black,line width=0.9pt,solid},
	{blue!85!black,line width=0.9pt,solid}
	}
\pgfplotsset{
	every axis/.append style={
		font=\scriptsize,
		width=\columnwidth,
		height=.8\columnwidth,
		legend style={at={(1.02,1)}, anchor={north west}},
		cycle list name=defaultcolorlist,
	},
	every axis title/.style={
		at={(0.5,1)},
		below
	},
}

\begin{document}

\title{Variance Reduction with Array-RQMC for Tau-Leaping Simulation of 
 Stochastic Biological and Chemical Reaction Networks} 
\titlerunning{Array-RQMC for Tau-Leaping Simulation}

\author{Florian Puchhammer \and Amal Ben Abdellah  \and Pierre L'Ecuyer}

\institute{F. Puchhammer \at
  Basque Center for Applied Mathematics, 
  Alameda de Mazarredo 14, 48009 Bilbao, Basque Country, Spain; and
  DIRO, 
	{Universit{\'e} de Montr{\'e}al},
	C.P.\ 6128, Succ.\ Centre-Ville, Montr\'eal, H3C 3J7, Canada \\
  \email{fpuchhammer@bcamath.org}  
 \and
  A. Ben Abdellah \at
  DIRO, 
	{Universit{\'e} de Montr{\'e}al},
	C.P.\ 6128, Succ.\ Centre-Ville, Montr\'eal, H3C 3J7, Canada \\
  \email{amal.ben.abdellah@umontreal.ca}           
 \and
  P. L'Ecuyer \at
  DIRO, 
	{Universit{\'e} de Montr{\'e}al},
	C.P.\ 6128, Succ.\ Centre-Ville, Montr\'eal, H3C 3J7, Canada \\
  \email{lecuyer@iro.umontreal.ca}           
}

\date{\today}

\maketitle 

\begin{abstract}
We explore the use of Array-RQMC, a randomized quasi-Monte Carlo method designed
for the simulation of Markov chains, to reduce the variance when simulating 
stochastic biological or chemical reaction networks with $\tau$-leaping.
The task is to estimate the expectation of a function of molecule copy numbers at a
given future time $T$ by the sample average over $n$ sample paths, and the goal is to reduce 
the variance of this sample-average estimator.
We find that when the method is properly applied, variance reductions by factors
in the thousands can be obtained.
These factors are much larger than those observed previously by other authors
who tried RQMC methods for the same examples.
Array-RQMC simulates an array of realizations of the Markov chain and requires
a sorting function to reorder these chains according to their states, after each step.
The choice of sorting function is a key ingredient for the efficiency of the method,
\new{although in our experiments, Array-RQMC was never worse than ordinary Monte Carlo,
regardless of the sorting method.}
The expected number of reactions of each type per step also has an impact on
the efficiency gain.
%
\keywords{Chemical reaction networks \and stochastic biological systems \and variance reduction 
  \and quasi-Monte Carlo \and Array-RQMC \and tau-leaping \and continuous-time Markov chains \and Gillespie}
\end{abstract}

\section{Introduction}

We consider systems of chemical species whose molecule numbers dynamically change over time as 
the molecules react via a set of predefined chemical equations. 
The evolution of such systems is typically modeled by a \emph{continuous-time Markov chain} (CTMC) 
\citep{sGIL77a,pAND91a,pAND11a} whose state is a vector that gives the number of copies of each species.
Each transition (or jump) of the CTMC corresponds to the occurrence of one reaction,
and the occurrence rate of each potential reaction (also called the \emph{reaction propensity}) 
is a function of the state of the chain.  
The probability that any given reaction is the next one that will occur is proportional to its propensity
and the time until the next reaction has an exponential distribution whose rate is the sum 
of these propensities. 
The \emph{stochastic simulation algorithm} (SSA) of \citet{sGIL77a} simulates the 
successive transitions of this CTMC one by one, by generating the exponential time 
until the next reaction and determining independently which reaction it is.
This method is exact (there is no bias).
But when the number of molecules is large, simulating all the reactions one by one
is often too slow, because their frequency is too high.
One popular alternative is to approximate the CTMC by a \emph{discrete-time Markov chain} (DTMC),
as follows.  Fix a time interval $\tau > 0$.  Under the simplifying assumption that the 
rates of the different reactions do not change during the next $\tau$ units of time, 
the numbers of occurrences for each type of reaction are independent Poisson random variables
with means that are $\tau$ times the occurrence rates (or propensities) of these reactions.
Each step (or transition) of the DTMC corresponds to $\tau$ units of time for the CTMC.
This DTMC can be simulated by generating a vector of independent Poisson random variables
at each step, and updating the state to reflect all the reactions that occurred during this
time interval.  In the setting of chemical reaction networks, this approach is the 
\emph{$\tau$-leaping method} of \cite{sGIL01a}, and it is widely used in practice.
This is the method we consider in this paper.

There are several other approximation methods, some of them leading to simpler and faster 
simulations, but the error and/or bias can also be more significant \citep{sGIL00a,sHIG08a}.
One simple approach uses a 
fluid approximation in which the copy numbers are assumed to take real values that vary 
in time according to a system of deterministic differential equations called the 
\emph{reaction rate equations} which can be simulated numerically \citep{sGIL00a,sHIG08a}.
This type of deterministic model is the primary tool in the field of system dynamics, 
and it is widely used in many areas.  
It corresponds to chemical kinetics equations found in textbooks. 
\hflorian{We should keep in mind that we could use this for sorting, though.}%
\hpierre{Yes but it would be distracting to talk about it here.}%
But this model ignores randomness, so it cannot capture the stochastic variations 
observed in experiments with real systems \citep{sBEE19a}.
Noise can be introduced via a \emph{stochastic differential equations} model,
which amounts essentially to approximate the Poisson distribution by a normal \red{distribution},
and the denumerable-state CTMC by a continuous-state process.
This leads to the \emph{chemical Langevin equation} \citep{sGIL00a,sBEE19a},
which can be simulated efficiently by standard methods for stochastic differential equations
\citep{mKLO92a} and may provide a reasonable approximation, \red{typically} when the number of molecules 
of each type is very large, but can otherwise suffer from bias.

The purpose of the stochastic simulations with $\tau$-leaping could be for example
to estimate the probability distribution of the state at a given time $t$, 
or the probability that the state is in a given subset, 
\red{or more generally the expectation of some function of the state, at time $t$.}  
The simulations are usually done via Monte Carlo (MC) sampling, using a random number generator 
that provides a good imitation of independent uniform random variables over the interval $(0,1)$
\citep{rLEC12a}.  
\red{To estimate the expectation of a random variable such as a function of the 
state at a given time, standard MC uses the average over $n$ independent simulation samples.
The accuracy of this estimate is usually assessed by computing a confidence interval on the 
true value.  Since the width of the interval is proportional to the (estimated) standard deviation,
which is the square root of the variance of the sample average, it is of high interest to
find alternative estimators with the same expectation, similar computing costs, and a much smaller variance.
With standard MC, the variance and the standard deviation of the sample average
converge as $\cO(n^{-1})$ and $\cO(n^{-1/2})$, respectively, which is rather slow. 
That is, to obtain one more significant digit of accuracy, as measured by the confidence interval,
we need to multiply the number $n$ of simulations by 100.
We would like to improve on this.}

\emph{Randomized quasi-Monte Carlo} (RQMC) is an alternative sampling approach 
which under favorable conditions can improve this convergence rate of the variance
\red{to $\cO(n^{-\alpha+\epsilon})$ for any $\epsilon>0$, for some constant $\alpha$
that can often reach 2, and even larger values in special situations
\citep{vOWE97b,vLEC02a,vLEC09f,vLEC18a,vLEC20m}.}
\emph{Quasi-Monte Carlo} (QMC) replaces the $n$ independent vectors of uniform random
numbers that drive the simulations by $n$ \emph{deterministic} vectors with a sufficient
number of coordinates to simulate the system and which cover the space (the unit hypercube) 
more evenly than typical independent random points \citep{rNIE92b,rDIC10a}.
RQMC randomizes these points in a way that each individual point becomes a vector of
independent uniform random numbers, while at the same time the set of points as a whole
retains its structure and high uniformity.  
\red{A point set that satisfies these two \emph{RQMC conditions} can provide 
an unbiased estimator with lower variance than Monte Carlo.} 

\red{RQMC also has} important limitations.
Firstly, the \red{$\cO(n^{-\alpha+\epsilon})$} convergence rates
are proved only under conditions that the integrand is a smooth function of the uniforms, 
whereas when simulating the CTMC considered here, the sequence of states that are visited 
is discontinuous in the underlying uniform random variates.
\hflorian{I think having integer-valued states is not a property of CTMCs in general, is it?}
\hpierre{The state space must be countable, so the chain evolves only by jumps.  
  The state does not have to be integer in general, but in our setting it is.
	However, this is not the key point.  I changed the formulation.}
Secondly, when the points are high-dimensional and some high-order interactions 
between the coordinates are important, the variance reduction is usually limited,
and this often happens when simulating the CTMCs that model reaction networks
via either direct SSA or $\tau$-leaping. Indeed, those simulations require at least one or two
random numbers per step of the chain, the number of steps can be very large in real applications,
so the dimension of the points, which is the total number of random numbers that are required
to simulate one realization of the process, can be very large.
\hflorian{I fear that people who are not familiar with RQMC might not know that (and understand why) we 
have to use one point for one chain. Maybe we should mention that the dimensionality is proportional
to the number of steps.}%
\hpierre{I expanded the explanation.}
\cite{sBEE19a} investigated the performance of $\tau$-leaping combined with traditional RQMC 
and found that the gain from RQMC compared to MC was small. 
They mentioned the two \red{well-known} limitations above as possible explanations for this behavior. 

The \emph{Array-RQMC} algorithm \citep{vLEC06a,vLEC08a,vLEC09d} 
has been developed precisely to recapture the power of RQMC
when simulating Markov chains over a large number of steps, as in the problem considered here. 
The empirical variance under Array-RQMC has been observed to converge faster than 
under MC in several examples from various areas, sometimes at 
\red{a rate near $\cO(n^{-2})$ empirically},
even for some examples where the cost function was discontinuous 
\citep{vDEM05a,vLEC07b,vLEC08a,vLEC09d,vDIO10a,vLEC16b,vBEN19b}.
The faster convergence has also been proven theoretically under certain conditions \citep{vLEC08a}.

Our present work was motivated by \cite{sBEE19a} and our aim is to see how Array-RQMC 
can improve upon MC and classical RQMC, first for the same examples as in their paper,
\red{then for a few more elaborate cases.}
\hflorian{Just to clarify: we also consider other examples. 
 However, that does not need to be mentioned here.}%
\hpierre{Yes.}
\cite{sHEL08a} also experimented with Array-RQMC, in combination with uniformization of the 
CTMC and conditional Monte Carlo (CMC) based on the discrete-time conversion method of \cite{vFOX90a}.
Their goal was to estimate the probability distribution of the state at a fixed time $t > 0$.
In this setting, CMC alone provably reduces the variance.
Empirically, with CMC, they obtained variance reductions by factors of about 20 in one example
and 45 in another example.
With the combination of CMC with Array-RQMC, they observed variance reductions by a factor
of about 100 with $n = 10^5$ for both examples. 
Thus, Array-RQMC provides an additional gain on top of CMC, by a factor of about 2.5 to~5.
\hpierre{What sort was he using?  We should say it somewhere and suggest that he may have
  done better by using a better sort.}
\hflorian{Hellander mentions the following on page 5, bottom left: 
``The trajectories are stored in a $(nd)$ matrix which is sorted using a least significant digit radix sort where each column is sorted using PROXMAP sort.'' By the way, he also mentions that using information from the stoichiometric matrix and/or not considering all state variables in the sort might make simulations more efficient. But he did not pursue this question further.}
\hpierre{Ok, then it seems that adding this sentence not clarify much and would not really help us.
  So I propose to just leave it as it is.}%
\hflorian{Fine with me. Btw, a few lines earlier he also mentions ``first the trajectories are
sorted according to the first component, then according to the second, etc.'', so he sorts lexicographically. In his first example (Coupled Flows), the final molecule numbers of $S_1$ and $S_2$ are both smaller than 100, so it's basically a batch sort. Similarly for his second example.}%
\red{(The \emph{variance reduction factor} (VRF) for a given method is defined as the 
variance of the standard MC estimator divided by the variance with the given method, 
for the same sample size $n$.)}

In this paper, we show how to obtain much larger variance-reduction factors with Array-RQMC.
We do this in the same setting as \cite{sBEE19a}, where the $\tau$-leaping method is
used to estimate an expectation at a given time $t$.
\hflorian{Just to clarify: They only estimate the expected copy numbers of each species
and with Array-RQMC we only do it for one selected species. However, that does not need
to be mentioned here already.}%
\hpierre{Yes, I know. I prefer to keep it short here, to avoid distractions.
 However, from reading the other papers, I think what people are really interested
 in is the \emph{distribution} of the copy numbers, not their expectation.
 This is what Hellander (2008) is doing. We can nevertheless keep our current examples.}%
We find empirically that the combination of $\tau$-leaping with the Array-RQMC algorithm 
can bring not only a significant variance reduction, but also an improved convergence rate, 
compared with plain MC. 

The main idea of the Array-RQMC algorithm is to simulate $n$ copies 
\red{(or sample paths)} of the Markov chain 
in parallel, in a way that the empirical distribution of the chain's states at any given step 
is closer to the exact theoretical distribution at that step
than with ordinary MC.
To achieve this, at each step, the first few coordinates of the RQMC point set are designated to match 
the points to the states, and the remaining coordinates are used to advance the chains by one step. 
This matching can be interpreted as sorting the chains in some particular order,
to match the ordering of the RQMC points.  In the simple case where the state is one-dimensional,
it suffices to enumerate the points by increasing order of their first coordinate and sort the 
chains by increasing order of their state.
For higher-dimensional states, one possibility is to use some kind of multivariate sort
to order both the points and the states; 
we will describe some of these sorts in Section~\ref{sec:sorting-strategies}.
Another approach is to define an \emph{importance function}, which maps 
 the state to a one-dimensional representative value, and sort the chains by that value. 
The choice of mapping can have a significant impact on the performance.
If the mapping is fast to evaluate, this approach can reduce the computing time significantly,
because a one-dimensional sort is usually much faster to execute than a multivariate one.  
To preserve the power of Array-RQMC, on the other hand, the importance function must 
provide a good estimate (or forecast) 
of the expected future value or cost, given the state at which it is evaluated.  
For this, it must be tailored to the problem at hand.
A good tradeoff between simplicity and prediction accuracy is not always easy to achieve, 
but it is \red{an important} ingredient for the performance of Array-RQMC. 
As a proof of concept that this approach can work for reaction networks, 
we experiment with a very simple one-step look-ahead importance function, 
and we find that it works \red{reasonably well in our examples.} 
In our \red{numerical experiments}, 
this approach is often competitive with the best multivariate sorts.
\hpierre{I removed a sentence here, since it was questionable.}
\red{We do not mean that this simple heuristic choice of importance function is always good.
There are probably situations where a different heuristic would be needed.
Our main message is that for a wide range of examples, it is not too hard to design a
reasonably simple and effective importance function.
Multivariate batch sorts used in previous papers are also competitive in general and offer
the best performance in some cases.} 
We also discuss briefly how more elaborate importance functions could be defined.
\hpierre{Do we still do that?}
\hpierre{We first produce sample data by simulating several runs of the $\tau$-leaping DTMC
with plain MC, then fit simple models of parameterized importance functions to this data,
and use these estimated importance functions with Array-RQMC in a second stage.
In our experiments, this approach was always competitive with the best multivariate sorts
in terms of variance reduction, and the sorting times were shorter.}
\hflorian{Not sure if we shall actually compare runtimes or leave everything as vague as here.}
\hpierre{We will probably be asked to do it later by the reviewers.
  Then the speed of the sorts will matter more!}

The remainder is structured as follows. 
In Section~\ref{sec:tau} we recall the fixed-step $\tau$-leaping method for the simulation 
of well-mixed reaction networks in its simplest form. 
Section~\ref{sec:rqmc} gives a short review of RQMC, the main point set constructions, 
and the underlying theory.
In Section~\ref{sec:arrayRQMC}, we define the Array-RQMC method 
and discuss some of the most prominent multivariate sorting algorithms.  
In Section~\ref{sec:experiments}, we describe the methodology used for our experiments 
and provide numerical results, with a discussion.   A conclusion follows.

\section{The CTMC Model and the $\tau$-Leaping Algorithm for Reaction Networks }
\label{sec:tau}

We consider a system comprised of $ \ell \geq 1 $ types of chemical species $ S_{1}, \dots , S_{\ell} $ 
that can react via $d \geq 1 $ reaction types (or channels) $ R_{1}, \dots , R_{d} $.
We assume that the species are well-mixed within a volume that does not change over time 
and whose temperature remains constant. 
%
Each reaction  $ R_{k} $, $k=1,\dots,d$, can be written as
\begin{equation*}
\alpha_{1,k} S_{1} +\cdots+ \alpha_{\ell,k} S_{\ell}
   \xrightarrow{c_{k}}  \beta_{1,k} S_{1} +\cdots+ \beta_{\ell,k} S_{\ell}, 
	\quad \alpha_{i,k}, \beta_{i,k} \in \NN_0,
\end{equation*}
where $c_{k}>0 $ is the reaction rate constant for $R_k$. 
Let $\bX(t) = (X_1(t),... , X_{\ell}(t)) \in\NN_0^{\ell}$, 
where $X_{i}(t)$ is the copy number 
(i.e., the number of molecules) of type $S_{i}$ at time $t$, for $i=1,\dots,\ell$ and $0\leq t\leq T$.
The process $\{\bX(t),\, t\ge 0\}$ is modeled as a CTMC with fixed initial state 
$\bX(0) = \bx_0$ and for which each jump corresponds to the occurrence of one reaction.
The jump rate (or \emph{propensity function}) for reaction $R_k$ is a function $a_k$ of the 
current state; it is $a_k(\bx)$ when $\bX(t) = \bx$.
This means that for a small $\delta > 0$, reaction $R_k$ occurs exactly once during the 
time interval $(t, t+\delta]$ with probability $a_k(\bx) \delta + o(\delta)$ 
and occurs more than once with probability $o(\delta)$.   
When $R_k$ occurs, the state changes from $\bx$ to $\bx + \bzeta_{k}$,
where $\bzeta_{k} = (\beta_{1,k} - \alpha_{1,k},\dots , \beta_{\ell,k} - \alpha_{\ell,k})$ 
is the stoichiometric vector for $R_k$. 
The standard for $a_k(\bx)$, which we assume in our examples,
\red{is the \emph{mass action kinetics} model:}  $a_k(\bx) = c_k \mathcal{h}_k(\bx)$ where 
$\mathcal{h}_k(\bx) = \prod_{i=1}^\ell \binom{x_i}{\alpha_{i,k}}$ represents the number of ways 
of selecting the molecules for reaction $R_k$ when in state $\bx = (x_1,\dots,x_\ell)$ \citep{sHIG08a}.
When in state $\bx$, the time until the next reaction has an exponential distribution with 
rate $\lambda(\bx) = \sum_{k=1}^d a_k(\bx)$, the probability that this reaction is $R_k$
is $a_k(\bx) / \lambda(\bx)$, and these random variables are independent.
The SSA of \cite{sGIL77a} simulates this CTMC directly.
However, when a large number of reactions occur in the time interval of interest, 
the direct simulation approach may be too slow. 

\cite{sGIL01a} proposed the \emph{$\tau$-leaping} algorithm as a way to speed up the simulation.
This approach discretizes the time into intervals of length $\tau > 0$, 
and it generates directy the number of occurrences of each type of reaction in each such interval.
If $\bX(t) = \bx$ at the beginning of an interval, it is assumed (as an approximation) that 
the rate of each reaction $R_k$ remains equal to $a_k(\bx)$ during the entire interval 
$[t, t+\tau]$.  Under this simplifying assumption, the number $D_k$ of occurrences of $R_k$
during this time interval has a Poisson distribution with mean $a_k(\bx) \tau$,
and $D_1,\dots,D_d$ are independent. These $D_k$ can be simulated easily via the inversion method,
by generating independent uniform random numbers over $(0,1)$ and applying the inverse of the
cumulative distribution function (cdf) of the appropriate Poisson distribution \citep{rGIL16a}.
The simulated state at time $t+\tau$ is then
$
  \bx + \sum _{k=1}^{d}  D_{k} {\bzeta}_{k}.
$
Repeating this at each step gives an approximating \emph{discrete-time Markov chain} (DTMC) 
$\{\bX_j,\, j\ge 0\}$ defined by $\bX_0 = \bx_0$ and 
\begin{equation}
\label{eq:tau-leap}
  \bX_{j} ~=~ \bX_{j-1} + \sum _{k=1}^{d}  D_{j,k} {\bzeta}_{k}, 
	        ~=~ \bX_{j-1} + \sum _{k=1}^{d}  F^{-1}_{j,k}(U_{j,k}) {\bzeta}_{k} 
	        ~\eqdef~ \varphi(\bX_{j-1}, \bU_j),  
\end{equation}
where $D_{j,k} = F^{-1}_{j,k}(U_{j,k})$,
$F_{j,k}$ is the cdf of the Poisson distribution with mean $a_k(\bX_{j-1})\tau$, 
$\bU_j = (U_{j,1},\dots,U_{j,d})$,
and the $U_{j,k}$ are independent uniform random numbers over $(0,1)$,
for $k=1,\dots,d$ and $j\ge 1$.
If $\tau$ is small enough, $\bX_j$ has approximately the same distribution as $\bX(j\tau)$,
so this DTMC provides an approximate skeleton of a CTMC sample path.

This $\tau$-leaping approximation has some potential problems, because it introduces bias
which can propagate across successive steps, and this bias can be important 
if $\tau$ is not small enough.
It is also possible to obtain negative copy numbers, i.e., some coordinates of some $\bX_j$
taking negative values.  Adaptive strategies and modifications of the algorithm have
been designed to prevent or handle this; see, e.g., \cite{sAND08a,sAND12a,sBEE19a},
and the references given there. 
We do not discuss these techniques in this paper.
Our main goal is to explore how Array-RQMC can be effectively combined 
with $\tau$-leaping \red{with a fixed $\tau$}, and we keep the setting simple to avoid distractions. 
\red{Implementing Array-RQMC in an adaptive setting (with variable $\tau$) 
would be more complicated; we discuss it briefly at the end of Section~\ref{sec:sorting-strategies}.}
In our experiments, we took $\tau$ small enough so we did not observe negative copy numbers.

Following \cite{sBEE19a}, we suppose that the objective is to estimate 
$\mu = \EE[g(\bX(T))]$ for a given time $T > 0$ and some function $g : \NN_0^\ell \to \RR$.
\hflorian{At some point we need to state that \cite{sBEE19a} only estimated copy numbers and 
that this is also what we do. Perhaps in the methodology-section?}%
These authors only took a coordinate projection for $g$ (i.e., they 
only estimated expected copy numbers) in their examples, and we do the same
\red{for most of our examples}, but what we do applies easily to other choices of $g$.
\hflorian{I modified the above sentence a little bit. 
  I exchanged \emph{the identity} by \emph{a coordinate projection}}%
\red{In one of our examples, we take $g(\bx)$ as the indicator that $\bx$ belongs to a given set $A$,
so $\mu = \PP[\bX(T) \in A]$.  In another example, we also make experiments in which $g(\bx)$ 
is the square or the cube of one coordinate.}  
\hflorian{We have not included these examples yet! Details and results are in Section~1.2.2  
in my notes. Also, the referee asks about what happens when we estimate \emph{all} copy numbers
at once, with one and the same sort.}%
We take $\tau = T /s$ where $s$ is a positive
integer that represents the number of steps of the DTMC that will be simulated.
To estimate $\mu$ with $\tau$-leaping and MC, 
we simulate $n$ independent realizations of the DTMC via 
\begin{equation}
\label{eq:dtmc}
 \bX_{i,0} = \bx_0,\qquad 
 \bX_{i,j} = \varphi_{j}(\bX_{i,j-1},\bU_{i,j}) 
              \quad\mbox{ for } j=1,\ldots,s \mbox{ and } i=0,\dots,n-1,
\end{equation}
where the $\bU_{i,j}$'s are independent uniform random points over $(0,1)^{d}$.
The estimator is
\begin{equation}
\label{eq:tau-leap-average}
   \hat\mu_n = \frac{1}{n} \sum_{i=0}^{n-1} g(\bX_{i,s}).
\end{equation}
We know that $\EE[\hat\mu_n] = \EE[g(\bX_s)] \approx \EE[g(\bX(T))] = \mu$
(we do not look at the bias $\EE[g(\bX_s)] - \EE[g(\bX(T))]$ in this paper) 
and $\Var[\hat\mu_n] = \Var[g(\bX_s)] / n$.
 
To use classical RQMC instead of MC, we simply replace the independent random points by
a set of $n$ vectors $\bV_i = (\bU_{i,1},\ldots,\bU_{i,s})$, \red{$i=0,\dots,n-1$}, 
which form an RQMC point set in $sd$ dimensions, as did \cite{sBEE19a}.
\red{The estimator in (\ref{eq:tau-leap-average}) obtained with these
points $\bV_i$ is the RQMC estimator, denoted $\hat\mu_{n,\rqmc}$.
In Section~\ref{sec:rqmc}, we provide a short review of how RQMC point sets 
are constructed, some theory, and many references.}

\section{Randomized Quasi-Monte Carlo}
\label{sec:rqmc}

\hpierre{If we need to add a significant tutorial section on RQMC, I think it should go here,
 because the aim of Section~\ref{sec:rqmc-points} is mainly to tell what points we used for Array-RQMC.
 We should ask the editor.}
   
%
The two most popular QMC construction methods are lattice rules (usually of rank 1)
and digital nets (typically in base 2).
For a \emph{lattice rule of rank 1}, with $n$ points in $s$ dimensions, 
one selects a vector ${\ba} = (a_1,\dots,a_s)$ with coordinates in $\{1,\dots,n-1\}$, 
and such that each $a_j$ is relatively prime with $n$.
The QMC point set is
\[
 P_n = \left\{{\bu}_i = (i {\ba}/n) \bmod 1, \, i=0,\dots,n-1\right\}.
\]
This point set has a very regular lattice structure.  
It can be randomized via a random shift modulo 1: 
Generate one random vector $\bU$ uniformly distributed over the unit hypercube $[0,1)^s$ 
and add this $\bU$ to each $\bu_i$, modulo 1.  The resulting (random) point set 
$\tilde P_n = \{\bU_i = (\bu_i + \bU) \bmod 1, \, i=0,\dots,n-1\}$
is called a randomly-shifted lattice rule. 
The shift preserves the structure and the conditions for an RQMC point set 
mentioned in the introduction are satisfied.
For more details, pictures, etc., see \cite{vHIC98c,vLEC00b,vLEC12a,vSLO94a}.
The choice of $\ba$ is important; we will return to this.

%
A \emph{digital net in base 2} has $n = 2^k$ points for some integer $k$.
One selects an integer $w\ge k$ (often, $w=k$) and $s$ generating matrices ${\bC}_1,\dots, {\bC}_s$ 
of dimensions $w\times k$ and of rank $k$, with binary entries.
To define the point ${\bu}_i = (u_{i,1},\dots,u_{i,s})$, for $i=0,\dots,2^k-1$,
we first write $i = a_{i,0} + a_{i,1} 2 + \cdots + a_{i,k-1} 2^{k-1}$, 
and then for each $j$ we compute
\begin{eqnarray*}
  (u_{i,j,1}, \dots,  u_{i,j,w})^\tr  
    &:=& {\bC}_j \cdot (a_{i,0}, \cdots, a_{i,k-1})^\tr \bmod 2 \quad\mbox{ and }\quad 
  u_{i,j} = \sum_{\ell=1}^w u_{i,j,\ell} 2^{-\ell}.
\end{eqnarray*}
%
Here, the parameters to select are the elements of the matrices ${\bC}_j$.
A popular way to construct them is to take $\bC_1$ as the reflected $k\times k$ identity matrix,
which gives $u_{i,1} = i/n$ for the first coordinate, then take for the other coordinates the 
generating matrices for the \emph{Sobol' sequence} \citep{rSOB67a,sLEM09a}.
These are upper triangular invertible $k\times k$ matrices constructed by specific rules,
but the bits above the diagonal in the first few columns can be selected arbitrarily, 
and their values have an impact on the uniformity of the higher-dimensional projections of $P_n$.
General-purpose choices are proposed in \cite{rJOE08a} and \cite{iLEM04a}.
Custom constructions can also be made by giving arbitrary weights to the different projections, 
using the LatNet Builder software \citep{vLEC20m}.

Applying a random shift modulo 1 to a digital net in base 2 does not preserve its 
digital net structure, but a random digital shift does and satisfies the RQMC conditions.
It consists in generating a single random vector $\bU$ uniformly over $[0,1)^s$, 
and doing a bitwise exclusive-or of all the bits of $\bU$
with the corresponding bits of each $\bu_i$ to obtain the randomized points $\bU_i$
\citep{vLEC02a,rDIC10a,vLEC18a}.
A much more elaborate and costly randomization method for digital nets is the 
\emph{nested uniform scramble} (NUS) of \cite{vOWE97a,vOWE97b}, 
described in many places including \cite{rDIC10a,vLEC18a}.
NUS became popular because \cite{vOWE97b} proved for known point set constructions that
for a sufficiently smooth $f$, the RQMC variance with NUS converges as ${\cO}(n^{-3+\epsilon})$
for any $\epsilon > 0$.  The same fast convergence rate was later proved by \cite{vHIC01a}
for a less expensive randomization which consists of a \emph{linear matrix scramble} (LMS) 
followed by a random digital shift (LMS+shift), as proposed by \cite{rMAT98c}. 
The LMS generates a random non-singular lower-triangular $w\times w$ binary matrix ${\bL}_j$
and replaces ${\bC}_j$ by ${\bL}_j {\bC}_j \? \bmod 2$, for each coordinate $j$.

For both the lattice rules and the digital nets, each one-dimensional projection of $P_n$ 
(truncated to its first $k$ digits in the case of the digital net) 
is $\{0, 1/n, \dots, (n-1)/n\}$. Thus, by looking at any one coordinate at a time,
the points cover the unit interval $[0,1)$ very evenly, which is already a good start.
Beyond this, the quality of $P_n$ must be measured by assessing the uniformity of its 
higher-dimensional projections, while giving more weights to the projections deemed more important.
This is usually done by working in a Hilbert space of functions $f : [0,1)^s \to\RR$.
The idea is to define a functional ANOVA decomposition of $f$ as 
$f = \sum_{\fraku\subseteq\{1,2,\dots,s\}} f_{\fraku}$ where $f_{\fraku}$ depends only
on the coordinates of $\bu$ whose indexes are in the set $\fraku$, and 
$\Var[f(\bU)] = \sum_{\fraku\subseteq\{1,2,\dots,s\}} \Var[f_{\fraku}(\bU)]$.
The important projections are those for which $\Var[f_{\fraku}(\bU)]$ is large.
The variance of the RQMC average
\begin{equation}
  \hat\mu_{n,\rqmc} = \frac{1}{n} \sum_{i=0}^{n-1} f({\bU}_i)   \label{eq:aver-qmc}
\end{equation}
is then bounded by a product of two terms, one that depends only on the point set $P_n$
and the other that depends only on the integrand $f$:
\begin{equation}
  \Var[\hat\mu_{n,\rqmc}] \le \cD^2(P_n) \cV^2(f),       \label{eq:HK}
\end{equation}
where
\begin{equation}
\label{eq:weightedVariation}
  \cV^2(f) = \sum_{\emptyset\not=\fraku\subseteq\{1,2,\dots,s\}}
	           \gamma_{\fraku}^{-2} \cV^2(f_{\fraku}) \
\end{equation}
and 
\begin{equation}
\label{eq:weightedFOM}
	\cD^2(P_n) = \sum_{\emptyset\not=\fraku\subseteq\{1,2,\dots,s\}}
	             \gamma_{\fraku}^{2} \cD^2_{\fraku}(P_n),
\end{equation}
where the $\gamma_{\fraku} \in\RR^+$ are weights assigned to the subsets $\fraku$ of coordinates, 
$\cV(f_{\fraku})$ measures the \emph{variation} of $f_{\fraku}$, and 
$\cD_{\fraku}(P_n)$ measures the \emph{discrepancy} (or non-uniformity) 
of the projection of $P_n$ over the subset $\fraku$ of coordinates.
For a function $f$ with finite variation $\cV(f)$, the RQMC variance in (\ref{eq:HK}) 
converges at the same rate as $\cD^2(P_n)$, so the goal becomes to construct point sets $P_n$
for which $\cD^2(P_n)$ converges to 0 as fast as possible when $n\to\infty$.
For the details, see \cite{rDIC10a,vLEC09f,vLEC20m,vOWE98a},
and references given there.

The decomposition in (\ref{eq:HK}) depends on the choice of function space.
The most classical version is the standard Koksma--Hlawka inequality, in which $\cD(P_n)$
is the star discrepancy of $P_n$ and $\cV(f)$ is the variation in the sense of Hardy and Krause
\citep{rNIE92b}.
\hflorian{Perhaps use ``function space'' instead of ``Hilbert space'' in the previous sentence. Otherwise,
people might think that the classical Koksma--Hlawka inequality is based on Hilbert spaces, which it isn't
as far as I know.}%
\hflorian{Koksma--Hlawka with a long hyphen ({-}{-}) or a short one (-)? 
 I think in the KDE paper they corrected it to a long one.}%
However, the star discrepancy is much too hard to compute to be used as a practical selection criterion.
Nowadays, one prefers to construct Hilbert spaces for which $\cD_{\fraku}(P_n)$ 
can be computed efficiently for the type of point set construction of interest.  
For example, for a randomly-shifted lattice rule with point set $P_n$, the most popular 
measure (called $\cP_\alpha$) has 
\begin{equation}
\label{eq:FOMfraku}
	\cD^2_{\fraku}(P_n) = \frac{1}{n} \sum_{i=0}^{n-1} \prod_{j\in\fraku} \phi(u_{i,j}),
\end{equation}
where $\phi(u_{i,j}) = -(-4\pi^2)^{\alpha/2} B_{\alpha}(u_{i,j}) / \alpha!$ 
for an even integer $\alpha \ge 2$ and $B_{\alpha}$ denotes the Bernoulli polynomial of degree $\alpha$.   
\hflorian{The symbol $B_{\alpha}$ had not been introduced.}%
For this measure, it is known how to construct lattice point sets $P_n$ such that
$\cD^2(P_n) = \cO(n^{-\alpha+\epsilon})$ for any $\epsilon > 0$, for any $s$ and finite weights
\citep{vDIC06a,vSIN12a,vLEC16a}.
It is also known that for periodic continuous functions whose mixed partial derivatives 
up to order ${\alpha/2}$ are square integrable, the corresponding variation $\cV(f)$ is finite.
The periodicity condition means that if $\bu$ has one coordinate $u_j$ at 0
and we replace the value of $u_j$ by 1 (or the limit as $u_j\to 1$), then $f(\bu)$ remains the same.
When $f$ is continuous but not periodic, we can easily transform it into an equivalent periodic 
function by applying a one-dimensional baker (or tent) transformation separately for each coordinate.
This transformation stretches each coordinate of each (randomized) point by a factor of 2,
then folds back the values by replacing $u$ with $2-u$ when $u > 1$.
This is equivalent to compressing the function horizontally by a factor of 2 and making a mirror
copy on the other half, which produces a periodic continuous function whose integral is the same 
as the original one. This can improve the convergence rate, as proved by \citep{vHIC02a},
and thus provide huge variance reductions in some cases.
On the other hand, it also increases the variation of the integrand, so it may increase the variance
(moderately) in other cases.

Similar theory and discrepancy measures have been developed for digital nets 
with random digital shifts and also for other types of scrambles such as NUS and LMS+shift.
The discrepancies that are practically computable usually have the same form as in 
(\ref{eq:weightedFOM}) and (\ref{eq:FOMfraku}), with different definitions of $\phi$.
For the details, see \cite{rDIC10a,vLEC20m}, and the references given there.

\color{black}

\section{Array-RQMC to Simulate the DTMC}
\label{sec:arrayRQMC}

\subsection{The Array-RQMC Algorithm}
\label{sec:arrayRQMC-algo}
\hpierre{In the previous version, the notation in this section did not match the notation
 of the previous section, as if the two sections were kind of extracted directly from two 
 different papers. The notations should match, and I think here we want to explain how
 Array-RQMC works for the problem considered in this paper.  General descriptions of 
 Array-RQMC are already available elsewhere.  So I rewrote the section in that direction.
 The index $i$ was starting sometimes at 0, sometimes at 1. It should always start at 0.}%
We now explain how to apply Array-RQMC to simulate the DTMC via (\ref{eq:dtmc}) and estimate 
$\EE[g(\bX_s)] \approx \mu$ again with (\ref{eq:tau-leap-average}), but with a different 
sampling strategy for the random numbers.  The algorithm 
simulates the $n$ sample paths of the DTMC in parallel, using an $(l+d)$-dimensional 
\hpierre{I have replaced $K$ by $d$ everywhere in the paper, to avoid taking
 a lowercase letter for one dimension and a capital letter for the other (this looked inconsistent).}%
RQMC point set to advance all the chains by one step at a time, for some $l \in\{1,\dots, \ell\}$.
The first $l$ coordinates of the points are used to make a one-to-one pairing 
between the chains and the points, and the other $d$ coordinates are used to advance the chains.  
When $l < \ell$, one must first define a \emph{dimension-reduction mapping} $h : \NN_0^\ell \to \RR^l$ 
whose aim is to extract the most important features from the state and summarize them in a 
lower-dimensional vector which is used for the sort.  
For $l=1$, the mapping $h$ has been called an \emph{importance function} or \emph{sorting function}
\citep{vLEC06a,vLEC07b}.
At each step, both the RQMC points and the chains are ordered using the same $l$-dimensional sort. 
Different types of sorts are discussed in Section~\ref{sec:sorting-strategies}.

Specifically, we select a deterministic low-discrepancy (QMC) point set of the form
$Q_{n}=\{(\bw_{i},\bu_{i}),\, i=0,\dots,n-1\}$, 
with $\bw_{i}\in [0,1)^{l}$ and $\bu_{i}\in[0,1)^{d}$, 
whose points are already sorted with respect to their first $l$ coordinates 
with the multivariate sort that we have selected. 
At each step $j$, we randomize the last $d$ coordinates of the points of $Q_{n}$
to obtain the RQMC point set
\begin{equation}
\label{eqn:Qnj}
   \tilde Q_{n,j}=\{(\bw_{i},\bU_{i,j}):\, i=0,\dots,n-1\},
\end{equation}
in which each $\bU_{i,j}$ is uniformly distributed in $[0,1)^{d}$. 
We also sort the $n$ states $\bX_{0,j-1}, \ldots,\? \bX_{n-1,j-1}$ based on their values of
$h(\bX_{0,j-1}),\ldots,h(\bX_{n-1,j-1})$,
using the same sorting algorithm as for the QMC points, and
let $\pi_j$ denote the permutation of the indices $\{0,1,\dots,n-1\}$ 
implicitly defined by this reordering.
Then the $n$ chains advance to step $j$ via 
\[
  \bX_{i,j} = \varphi(\bX_{\pi_j(i),j-1}, \bU_{i,j}), \qquad   i=0,\ldots,n-1.
\]
It is also possible to use a different sorting method at each step $j$, 
in which case the QMC points must be sorted differently as well, 
so this is usually not convenient.

At the end, one computes $\hat\mu_n$ as in (\ref{eq:tau-leap-average}), which \red{gives}
an unbiased estimator of $\EE[g(\bX_s)]$. 
The main goal of this procedure is for the empirical distribution 
of the states $\bX_{0,j},\ldots,\bX_{n-1,j}$ to better approximate the theoretical 
distribution of $\bX_{j}$ at each step $j$, than if the chains were simulated 
independently with standard MC, and as a result reduce the variance of $\hat\mu_n$.
\hpierre{Here we should add an intuitive explanation to justify this approach.
  Perhaps something similar to \cite{vLEC16b}. Reformulate what follows.}%
\red{The following heuristic argument gives insight on why it works.
To simplify, suppose that $l=\ell$ and the state $\bX_j$ has the uniform distribution over $[0,1)^\ell$,
for each $j$. This can be obtained conceptually by a monotone change of variable
(which does not have to be known explicitly).   
At step $j$, for any function $g_j : [0,1)^\ell\to\RR$, the algorithm estimates 
\[
  \EE[g_j(\bX_j)] = \EE[g_j(\varphi(\bX_{j-1},{\bU}))] 
	= \int_{[0,1)^{\ell+d}} g_j(\varphi({\bx},{\bu})) d{\bx} d{\bu}
\]
by the average
\[
       \frac{1}{n} \sum_{i=0}^{n-1} g_j(\bX_{i,j}) 
    =  \frac{1}{n} \sum_{i=0}^{n-1} g_j(\varphi(\bX_{i,j-1}, {\bU}_{i,j})).
\]
This is essentially an RQMC estimate with the point set 
$\cQ_{n,j} = \{(\bX_{i,j-1}, {\bU}_{i,j}),\, 0\le i < n\}$.
We would like $\cQ_{n,j}$ to be highly uniform over $[0,1)^{\ell+d}$ but we cannot
really choose the $\bX_{i,j-1}$'s in these points, since they depend on the simulation.
What we do instead is select the RQMC point set $\tilde Q_{n,j}$ defined above
and reorder the states $\bX_{i,j-1}$ in a way that 
$\bX_{i,j-1}$ is close to $\bw_i$ for each $i$.
This is the role of the sorting step.
For a more detailed} 
theoretical analysis and empirical evidence, see for example \cite{vLEC08a,vLEC09d,vLEC16b}.
To estimate the variance of this Array-RQMC estimator, one can repeat the entire procedure
$m$ times, with independent randomizations of the points, 
and take the empirical variance of the $m$ realizations of $\hat\mu_n$ as an 
unbiased estimator for $\Var[\hat\mu_{n}]$, \red{as with classical RQMC}.
This Array-RQMC procedure is stated in Algorithm \ref{alg:arrayalgorithm}.

\new{One may wonder how easily this algorithm can be parallelized.
Of course, it is easy to run the $m$ independent replications in parallel,
although we usually prefer to use a large $n$ and small $m$ (e.g., 10 to 20) to benefit from the 
improved convergence rate.  
On the other hand, running a single Array-RQMC replication (with $n$ chains) in parallel 
(e.g., on a GPU card) is more complicated, because of the sorting at each step.
This aspect has to be further investigated.
Parallelization would not change the variance, but may increase the speed.}
\hflorian{Due to the improved convergence rate in $n$, one would usually prefer to choose $n$ larger
than $m$ and thus to parallelize the Array-RQMC algorithm differently. The dependence between 
the $n$ chains and the sorting step in particular, make this task challenging
and it is currently investigated how to do that. The gains of Array-RQMC over MC in terms
of variance are often significant, even for moderate $n$, so that its application is usually worthwhile.}

\begin{algorithm}[H]
	\caption{Array-RQMC Algorithm}
	\label{alg:arrayalgorithm}
	\begin{algorithmic}[1]
		\State  $\bX_{i,0} \leftarrow \bx_{0}  $  for $i=0,...,n-1$; 
		\For{$ j=1,2,...,s $}
		  \State Sort the states $\bX_{0,j-1},\dots, \bX_{n-1,j-1}$ by their values of $h(\bX_{i,j-1})$,
		  \State \q using the selected sort, and let $\pi_j$ be the corresponding permutation;
			\State \q \red{// We assume that the points are sorted in the same way by their first $l$ coordinates;}
		  \State Randomize afresh the last $d$ coordinates of the RQMC points, 
		       $\textbf{U}_{0,j},...,\textbf{U}_{n-1,j}$; 
		  \For{$ i = 0,1,\dots,n-1 $}   		 
		     \State $\bX_{i,j} = \varphi(\bX_{\pi_j(i),j-1},\bU_{i,j})$ ;  		
		  \EndFor
		\EndFor
		\State Return the estimator        
		  $\hat\mu_n = (1/n) \sum_{i=0}^{n-1} g(\bX_{i,s})$.
	\end{algorithmic}
\end{algorithm}

\new{A user may want to fix an accuracy target and increase $n$ or $m$ 
adaptively until the target is reached.  
This is easily implemented by increasing $m$ for a fixed $n$, but again we prefer increasing $n$.
Then, we have to re-run the algorithm with the larger $n$, because the result of the sort 
is different with the larger $n$. A reasonable strategy would be to first run the algorithm say 
with $(n,m) = (n_1, m_1)$, and estimate from that a pair $(n,m) = (n_2, m_2)$ that would meet 
the target accuracy. If we find that it would suffice to increase $m$ by a factor of no more than
2 or 3 with the same $n$, we just do that.  
Otherwise, we re-run the algorithm with the larger $n=n_2$ deemed sufficient, 
and we can use a linear combination of the two estimators.}

\hflorian{To select $(n,m)$ in practice, one can try a pair
$(n_1,m_1)$ that fits the computational budget. If the results of this simulation are deemed insufficent,
one can use them to decide how to proceed. If a larger $m$ is needed, then simply run the additional simulations with the same $n_1$. If $n_1$ was too small, then we need to re-run the entire simulation with
 $n_2>n_1$, as the results of the sorting steps will be different. But one can use the earlier results for variance reduction, by combining the two estimators via a convex combination, which is equivalent to using control variates \cite[Problem 2.3.9]{sBRA87a}.}

\subsection{Sorting Strategies}
\label{sec:sorting-strategies}

In the special case where $l=1$, the RQMC points are sorted by their first coordinate
and the states $\bX_{i,j-1}$ are simply sorted by their value of $h(\bX_{i,j-1})$, 
in increasing order.  In this case, one would typically have $\bw_{i}= i/n$ and the 
points are already sorted by construction (this is true for all the point sets used in this paper).

When $\ell>1$, sorting for good pairing is less obvious. 
\red{One multivariate sort that gave good results for other applications is the 
\emph{batch sort}, defined as follows \citep{vLEC98a,vELH08a,vLEC09d,vLEC16b}. 
We factor $n \approx n' = n_{1}n_{2}\cdots n_{L}$ with $1\le L\le d$ and $n \le n'$.
\hflorian{I don't think $n \le n'$ is true. Actually, in our examples it's usually
the opposite. E.g., choosing $n=2^{19}$ and batch exponents $\balpha=(1/2,1/2)$ we get
$n_1 n_2=19,575$, which is much smaller than $n$.}%
The approximation is because $n$ is not always easy to factor; it can be a prime number for example.}
Each time we sort, we split the set of states into $ n_{1} $ batches of size \red{approximately} $n/n_{1}$ 
such that the first coordinate of every state in one batch is smaller or equal to the 
first coordinate of every state in the next batch; 
then we further subdivide each batch into $ n_{2} $ batches of size \red{approximately} $n/(n_{1}n_{2})$ 
in the same way but now according to the second coordinate of the states. 
This procedure is repeated $L$ times in total.  
\red{In practice, $L$ should rarely exceed 3.
If $n < n'$, some batches (the last ones) will contain fewer states. 
Here, the required dimension for $\bw_i$ in (\ref{eqn:Qnj}) is $l = L$.
In our experiments, when we apply the method for several values of $n$, 
the choices of $n_{1}, n_{2}, \dots, n_{L}$ must depend on $n$.
What we do is select a vector of positive exponents $\balpha = (\alpha_1,\dots,\alpha_L)$
such that $\alpha_1 + \cdots + \alpha_L = 1$, called the \emph{batch exponents}, and put 
$n_j=\lceil n^{\alpha_j}\rceil$ for all $j$.
We will report those batch exponents.} 

Another way of sorting is to map the states to the $\ell$-dimensional unit 
hypercube $[0,1)^{\ell}$, so we can assume that the state space is now 
$[0,1)^{\ell}$ instead of $\NN_0^\ell$,  
and then use a discretized version of a space filling curve for this hypercube.
The hypercube is partitioned into a grid of small subcubes so that the event that two states
fall in the same small subcube has a very small probability, then the states are sorted in
the order that their subcubes are visited by the curve (those in the same subcube can
be ordered arbitrarily).  With this, we use $(d+1)$-dimensional RQMC points sorted by their
first coordinate.  This approach is in fact an implicit way to map the states 
to the one-dimensional real line, and then use a one-dimensional sort (with $l=1$).  
This has been suggested in particular with a Z-curve \citep{vWAC08a} and with a Hilbert curve 
\citep{tGER15a}.  We call the latter a \emph{Hilbert curve sort}.
To map $\ell$-dimensional states to $[0,1)^{\ell}$, \cite{tGER15a} suggest applying 
a rescaled logistic transformation 
$\Psi(x_j)= 1/(1+\exp[-(x_j - \mu_j + 2\sigma_j)/(4\sigma_j)])$, $1\leq j\leq\ell$, 
to each coordinate. 
We estimated the means $\mu_j$ and the variances $\sigma_j^2$ of the copy numbers of each species 
at every step, from data obtained from preliminary experiments (pilot runs).
\red{We tried various numbers of pilot runs, from $2^4$ to $2^{19}$, and the results were not
significantly better with more pilot runs.  
This indicates that only very crude estimates are sufficient.}
\hflorian{Should we also mention the on-the-fly method (the referee suggests it), or only indicate 
in the response letter that we have tried it with no clear trend as to whether it does better or 
worse than using a few pilot runs.}
\hpierre{With how many pilot runs?   Mention the drawbacks of this approach.  
  Also, what would be a good transformation.   It is not our favorite....}%
\hflorian{We moved away from moving the standard normal CDF a while ago. I never actually knew
why would use it in the first place \ldots}
\hflorian{We \emph{can} state that ``this is \textbf{far} more than necessary'', if we want to stress
that. It would still fit the line.}

\hpierre{A variant that avoids the need for such a transformation is the 
\emph{Hilbert batch sort} \citep{vLEC16b}: One first applies a batch sort to partition 
$\RR^{\ell}$ into $n$ boxes, each containing exactly one of the states.
Then these boxes are associated with $n$ subcubes in $[0,1)^\ell$
and the states are enumerated in the order that the corresponding boxes are 
visited by the Hilbert curve.}

These multivariate sorts can be computationally expensive when $n$ is large.
For this reason, we made some efforts in this work to explore ways of defining importance functions 
$h: \NN_0^\ell\to\RR$ that can be computed quickly during the simulations and provide at the same time 
good representations for the value of a state.
An appropriate choice of $h$ is certainly problem-dependent and good ones have been 
constructed for some examples in other settings such as computational finance, 
queueing, and reliability \citep{vLEC07b,vLEC08a,vLEC16b,vBEN19b}.

We adopt the (partly heuristic) idea that at each step $j$, an ideal importance function $h_j$
should have the property that $h_j(\bx)$ is a good approximation of 
$\EE[g(\bX_s) \mid \bX_j = \bx]$ for all $\bx\in \NN_0^\ell$ and $j=1,\dots,s$ \citep{vLEC07b,vLEC09d}.
\hpierre{We must explain what is the rationale for this.  We should also check again the references
 just given (my papers) to see how we were explaining it.}%
\red{The rationale is that this conditional expectation can be seen as the ``value'' 
of the current state $\bx$, and can be taken as a ``summary statistic'' in place of the 
multidimensional state. In particular, two states with equal ``value'' can be considered equivalent.
To really implement this type of approximation, we need to construct a different $h_j$ for each $j$,
because the conditional expectation depends on $j$.}
We will call \red{this} a \emph{step-dependent importance function} (SDIF).
\red{One \emph{does not} need to use the expensive procedures 
that we now describe to be able to apply Array-RQMC for a given reaction network.  
Our goal is rather to see if these elaborate procedures are worthwhile, 
or if there are much simpler strategies that can perform almost as well.}
To see how well \red{a general SDIF} could perform, we made the following experiment
with each of the examples considered in Section~\ref{sec:experiments}.
First, we generated data by simulating the DTMC for $n = 2^{19}$ independent ``pilot'' sample paths,
and we collected the $n$ pairs $(\bX_{i,j}, g(\bX_{i,s}))$, $i=0,\dots,n-1$, for each $j$.
Then, our aim was to find a function $h_j : \NN_0^\ell \to \RR$ for which $h_j(\bX_{i,j})$ was 
a good predictor of $g(\bX_{i,s})$ conditional on $\bX_{i,j}$.  
For this, we selected a parameterized form of function $h_j$,
say $h_j(\btheta,\cdot)$, which depends on a parameter vector $\btheta$, 
and we estimated the best value of $\btheta$ by least-squares regression from the data.  
The general form that we explored for $h_j(\btheta,\bx)$ was a linear combination of 
polynomials in the coordinates of $\bx$, where $\btheta$ was the vector of coefficients
in the linear combination. The motivation for this choice is that the expected number of molecules
of a given type at the next step, given the current state, is an affine function of 
the expected number of reactions of each type that will occur at that step, and this expected 
number for reaction type $R_k$ is in turn linear in $a_k(\bx)$, which is a known polynomial in the 
coordinates of $\bx$.  

Let $\tilde h_j$ denote the functions $h_j$ estimated from data as just described, for each $j$.
These $\tilde h_j$ are noisy estimates, and since they are estimated separately across 
values of $j$, we can observe some random variation when looking at their sequence as a 
function of $j$.  To smooth out this variation, we tried fitting a
(least-squares) smoothing spline \citep{mDEB01a,mPOL93a} to this sequence of functions $\tilde h_j$ to
obtain a sequence of functions $h_j$, $j=1,...s$, that varies more smoothly across the step number $j$.  
This yields a \emph{smoothed SDIF}.
In our experiments, we never observed a large improvement by doing this, because with $n=2^{19}$
pilot simulations, the $\tilde h_j$ did not vary much already as a function of $j$.
But the smoothing might be worthwhile when the number $n$ of pilot simulations is smaller.

A cruder but less expensive strategy uses the same function $h_j = h$ for all $j$.
One \red{possibility} is to use $h_{s-1}$ at all steps,
\red{i.e., take 
\[
  h(\bx) = h_{s-1}(\bx) \eqdef \EE[g(\bX_s) \mid \bX_{s-1} = \bx] 
	       = \EE[g(\bX_1) \mid \bX_{0} = \bx].   
\]}
We had some success with this simple version, which we call the 
\emph{one-step look-ahead importance function} (OSLAIF).

In the special case where $g(\bx)$ is linear in $\bx$, say $g(\bx) = {\bf b}^\tr \bx$
where ${\bf b}^\tr$ is the transpose \red{of} a vector of coefficients, 
then \red{$h(\bx) = \EE[{\bf b}^\tr \bX_1 \mid \bX_{0} = \bx]$} 
is given by a polynomial in $\bx$, and one can obtain this polynomial exactly, since 
\red{from (\ref{eq:tau-leap}),}
\begin{eqnarray}
 \label{eq:econd-onestep}
  \EE[\bX_1 \mid \bX_{0} = \bx]
	&=& \bx + \sum_{k=1}^d \bzeta_k \EE[D_{1,k} \mid \bX_{0} = \bx] 
	~=~ \bx + \tau \sum_{k=1}^d \bzeta_k a_k(\bx),
\end{eqnarray}
which is a vector of polynomials in $\bx$ that are easy to calculate.
This includes the case of $g(\bx) = x_i$, the number of molecules of species $i$, 
which occurs in our examples.

Extending this to more than one step can be more difficult when the $a_k$ are nonlinear.
One can write 
\begin{eqnarray*}
  \EE[\bX_2 \mid \bX_{0} = \bx]
	&=& \bx + \tau \sum_{k=1}^d \bzeta_k \left[a_k(\bx) + \EE[a_k(\bX_1) \mid \bX_{0} = \bx]\right],
\end{eqnarray*}
but when $a_k$ is nonlinear, the quantity in the last expectation is a nonlinear function 
of a random vector.  
Extending to more steps leads to even more complicated embedded conditional expectations.
This motivated us to try just the OSLAIF rule as a heuristic, and we 
\red{already obtained satisfactory} results with that.
Specific illustrations are given in Section~\ref{sec:experiments}.
\red{This OSLAIF heuristic also works when $g$ is nonlinear, e.g., for higher moments or 
for an indicator function, as long as we can compute or approximate the expectation.  
We will give examples of that in Section~\ref{sec:experiments}.}
\hpierre{Not sure if this simple rule is always reasonable. 
 This may work very well in some small examples,
 but very poorly in other cases.  What we would really want to approximate is 
 $\EE[g(\bX_s) \mid \bX_{j} = \bx]$ for each $j$, and there could be an important difference 
 between different values of $j$, in particular when $s-j$ is large.}
\hflorian{I think it cannot be reasonable for many examples. It should work well if the dynamics
 are more or less the same at every time $t$. Apparently (and surprisingly), this seems to be the
 case for our model problems. But just think of a system where the copy numbers oscillate a lot as
 time increases. Then one step simply cannot reflect the behavior of the entire trajectory.}
\hpierre{Yes, this is what I had in mind.}
 
\red{Using tau-leaping with a fixed $\tau$ goes along well with Array-RQMC 
because all the chains are then synchronized in time; they all advance by the same time step $\tau$
at each step and they all reach $T$ after the same number of steps.
This does not hold if we simulate the sample paths reaction by reaction, because the times between 
successive steps are then independent exponential random variables, so the number of steps is random.  
This lack of synchronization also occurs if the step size $\tau$ is variable and selected adaptively
for each sample path. In these cases, it could be a good idea to also include the current 
clock time $t$ in the list of state variables used for sorting.
Array-RQMC still applies when the number of steps is random and differs across the sample paths,
as explained in \cite{vLEC08a}, but the variance reduction is typically more modest in that case.
This should be explored in future research. }

\subsection{RQMC Point Sets}
\label{sec:rqmc-points}

\hpierre{We have been asked to expand our explanations of RQMC points, 
   and justify the different choices of parameters. }
\red{We report results for the following point sets in this paper
(the short names in parentheses are used to identify them in the next section):
(1) a randomly-shifted rank-1 lattice rule (Lat+s); 
(2) a Lat+s with the baker's transformation applied to the points after the shift (Lat+s+b);
(3) a Sobol' net with a left random matrix scramble followed by a random digital shift (Sob+LMS).
In our experiments, we also tried Sobol' nets with the nested uniform scramble 
of \cite{vOWE97b} (Sob+NUS), but the variance was about the same for Sob+LMS
and the computing times were significantly longer, so the \eif{} was never better.
Therefore, we omit these results from the tables.}
All these methods are explained in Section~\ref{sec:rqmc}. 
They are all implemented in SSJ \citep{sLEC05a,iLEC16j}, \red{a general-purpose 
Java library for stochastic simulation which we used for all our experiments.
It provides the required RQMC tools and also implements the sorting methods discussed 
in Section~\ref{sec:sorting-strategies} 
(see the package \texttt{umontreal.ssj.util.sort} in SSJ).
For the one-dimensional sorts, these methods use the default quicksort implementation 
available in Java.
For classical RQMC, we used the point coordinates sequentially in time, and for each time step
 we used them in the same order as the reactions are numbered.
 For Array-RQMC, we used the first coordinates in the same order as they are used for the sort
 (e.g., for the batch sort), then the other coordinates in the same order as the reactions are numbered.
The MRG32k3a random number generator of \cite{rLEC99b} was used for MC and all the randomizations.}
\new{The Java code for our examples can be found in a repository available at 
\url{https://github.com/FlorianPuchhammer}.}

For the lattice rules, the parameters were found with the Lattice Builder tool \citep{vLEC16a},
using the weighted $\mathcal{P}_2$ criterion 
\red{defined via (\ref{eq:weightedFOM}) and (\ref{eq:FOMfraku}) 
with order dependent weights $\gamma_{\fraku}^2 = \rho^{|\fraku|}$ for $\rho = 0.6$.   
This choice is certainly not optimal, but it gave reasonably good results for all cases.
We tried other values of $\rho$ and smaller values, for which the weight 
decreases much faster with the dimension, gave better results for some examples
(e.g., $\rho = 0.05$ for the example in Section~\ref{sec:schlogl}), 
but we nevertheless report the results for a single $\rho$, to show that this is already good enough.
For the Sobol' points, we used the parameters (direction numbers) from \cite{iLEM04a}
(our preference) for Array-RQMC.  But these parameters are given only for up to 360 dimensions,
which is often not enough for classical RQMC, so we used the table from \cite{rJOE08a}
in that case.}
%
\hflorian{I might downsize it to $n=2^{19}$ to be able to do something in meaningful time.}
\hpierre{If necessary, ok, but I would rather cut out on the items that take too much time,
 like the expensive sorts  and the expensive RQMC point sets, such as NUS. 
 Also make sure you do MC only once, not repeat 100 times!} 
\hflorian{Unfortunately, I had to downsize to $n=2^{19}$. At least for the first version of the paper. 
 I still need to update the document accordingly.
 I run the experiments for every point set individually in parallel. So, if one item 
 like a NUS or a split sort takes longer, it is not really dramatic. But if every item takes a lot
 of time, I might not have been able to finish them before I lose my cluster privileges.}
\hpierre{I believe that we should remove the Sobol'+NUS results from all the tables.
 This scrambling is practically never winning.  We can mention it briefly here, but remove it
 from the results.  I also thought about removing the baker's transformation, but then we would loose 
 something in the last panel of Table 1.}

\section{Numerical Illustrations}
\label{sec:experiments}

For our numerical illustrations, we use two low-dimensional examples taken from \cite{sBEE19a}, 
then a higher-dimensional example from \cite{sPAD16a}, \new{and one further example taken from
\cite{sKIM15a} to study the effect of Array-RQMC on quasi-steady state approximation}.
On these examples, we compare the performances of both classical RQMC and Array-RQMC 
in combination with $\tau$-leaping.

We repeated each Array-RQMC procedure $m=100$ times independently to estimate the RQMC variance
$\Var[\hat\mu_n]$ for $n=2^{13},\dots,2^{19}$.
We then fitted a model of the form  $\var[\hat\mu_n] \approx \kappa n^{-\beta}$ to these observations
by least-squares linear regression in log-log scale.
This gave an estimated convergence rate of $\cO(n^{-\hat\beta})$ for the variance, where $\hat\beta$
is the least-square estimate of $\beta$.  
We report $\hat\beta$ in our results.  
Ordinary MC gives $\beta=1$ \red{(exactly)}, so we can compare. 
We should keep in mind that the $\hat\beta$ are only noisy estimates 
and the linear model for the RQMC and Array-RQMC methods is only an approximation.
We also provide a few plots of $\Var[\hat\mu_n]$ as a function of $n$, in log-log scale,
to illustrate the typical behavior. 
The logs are all in base 2, because we use powers of 2 for $n$.

We computed the estimated \emph{variance reduction factor} (VRF) of Array-RQMC compared with MC,
which is defined as $\Var[g(\bX_{s})]/(n \Var[\hat\mu_n])$ where 
$\Var[g(\bX_{s})]$ is the MC variance for a single run, which was estimated separately
by making $n = 10^{6}$ independent runs.   
\hflorian{MC is run only once but with $10^{6}$ points. From this run, the 
 variance estimated, and then adjusted to the $n$ we are looking at. Doing MC as a RQMC 
 rule is only sometimes a good indicator to check, if the code is working properly.}%
\hpierre{Yes, but it can be much longer than the other RQMC point sets.}%
This is the variance per run for MC divided by the variance per run for Array-RQMC.
We call \vrf{} this value for $n=2^{19}$ and we report it in our results. 
\new{The VRF for other values of $n=2^{k}$ can be estimated via
$\mbox{\sc VRF}(k) \approx (\kappa_0 n^{-1}) / (\kappa n^{-\beta}) 
\approx \vrf{} / 2^{(\beta-1)(19-k)}$.}

We also computed an \emph{efficiency ratio} which measures the change in 
the work-normalized variance (the product of the estimator's variance by its computing cost).
It is the VRF multiplied by the CPU time required to compute $n$ realizations with MC
and divided by the CPU time to compute the RQMC or Array-RQMC estimator with the same $n$.
We call \eif{} its value for $n=2^{19}$ and we report it as well.
This measure takes into account both the gain in variance and the extra cost 
in CPU time which is required to sort the chains at each step of the Array-RQMC algorithm.
Note that using RQMC only is generally not slower than MC; \red{it is often} a bit faster.
\red{These \vrf{}'s and \eif{}'s should be understood as only providing noisy estimates, 
because the \vrf{}'s are only estimates based on $m=100$ replications, 
so there could be over 5\% error on these estimates. 
The true variances depend on the selected parameters for the generating vectors or matrices of the 
RQMC point sets, and the sorts. Also, points generation, randomizations, and the sorts are not necessarily 
implemented in the best possible way given the hardware.
\hflorian{I have just realized, that construction of the RQMC points is NOT included in the 
CPU time, only the randomization. This concerns both array-RQMC and RQMC. The current 
implementation would not allow that.}%
\hflorian{I found this quite hard to read. I made very slight alterations and inserted two periods. So please check if everything still has its intended meaning.}%
As an illustration, a batch sort with $n_1=n$ might run slightly faster than defining an 
importance function that uses only the first coordinate, even though the two are equivalent.
On the other hand, the gains we have obtained are sufficiently large to be convincing.
The VRF and EIF for smaller values of $n$ can be estimated by using \vrf{} and \eif{}
together with $\hat\beta$.
\hpierre{It may depend on which functions	were called in SSJ.  
 For instance, Korobov lattices can be generated much faster than 
 general rank-1 lattices. Also, the split sort has a very slow implementation.}%
For the timing comparisons, each RQMC or Array-RQMC experiment with one type of point set,
one type of sort, all values of $n$, and $m=100$, was executed as one job on a 
Lenovo NeXtScale nx360 M5 node with two Intel Xeon E5-2683 v4 cores at 2.1 GHz.}

\hpierre{Note that I have removed the split sort, Hilbert batch sort, and NUS everywhere.
  These were ading no value to the paper.}
\hflorian{Ok. I agree.}

\subsection{A Reversible Isomerization System }
 
We start with the same simple model of a \emph{reversible isomerization system}
as \cite{sBEE19a}.  There are two species, $ S_{1} $ and $ S_{2} $, 
and $d=2$ reaction channels with reaction rates $c_1 = 1 $ and $ c_2 = 10^{-4}$:
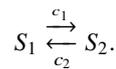
\begin{equation*}
   S_{1}  \stackrel{\xrightarrow{c_1}}{\xleftarrow[c_2]{}} S_{2}.
\end{equation*}
\hflorian{Just mention it once more: I am not sure if it dangerous to just sweep the relation
$\EE[\bX_t]=\bx_0$ for this parameter setting under the rug. B\&B mention that explicitly
for this example.}%
\red{There are initially} $X_1(0) = 10^2$ molecules of type $S_1$ and $X_2(0)=10^6$ molecules of type $S_2$.
Since the total number of molecules is constant over time, it suffices to know the number 
of molecules of the first type, $X_1(t)$, at any time $t$, so we can define the state of the 
CTMC as $\bX(t) = X_1(t)$ only.  
This gives $\ell=1$, and we only need a one-dimensional sort for Array-RQMC.
We also take $g(\bX(t)) = X_1(t)$.
With our choice of initial state,  
$\EE[X_1(t)] = 10^2$ for all $t >0$, so we already know the answer for this simple example.
There are two possible reactions, so $d=2$, and we therefore need RQMC points in 
$2s$ dimensions with classical RQMC and in $\ell+d =3$ dimensions with Array-RQMC.

\begin{table}[!htbp] 
 \centering
 \caption{Estimated rates $\hat\beta$, $\vrf$, and $\eif$, for the reversible isomerization example, for various choices of $(T,s,\tau)$.
	MC refers to ordinary MC, RQMC is classical RQMC with Sobol' points and LMS randomization, 
	and the other four rows are for Array-RQMC with different RQMC point sets. 
	``MC Var'' is $\Var[g(\bX_s)]$, the variance per run with MC.
	\red{For each case, the best value across the sampling methods is in bold.}}
\label{tab:revIso}
\small
 \begin{tabular}{l| crr| crr| crr }
\hline
 $(T,s,\tau)  \longrightarrow $ 
   & \multicolumn{3}{c|}{$(1.6,\,8,\,0.2)$}        & \multicolumn{3}{c|}{$(1.6,\,128,\,0.2/16)$} 
	 & \multicolumn{3}{c}{$(1.6,\,1024,\,0.2/128)$}  \\
\hline						
 MC $\Var$ &\multicolumn{3}{c|}{$107.8$}     &\multicolumn{3}{c|}{$96.6$} 
	 &\multicolumn{3}{c}{$96.0$} \\
\hline
Point sets  & $ \hat\beta $ & $\vrf$ & $\eif$ & $ \hat\beta $ & $\vrf$ & $\eif$
            & $ \hat\beta $ & $\vrf$ & $\eif$ \\
\hline
MC		    & 1.00	& 1      & 1       & 1.00	& 1        & 1     & 1.00	& 1	    & 1	 \\
RQMC        & 1.03	& 629	 & 1,493   & 1.08 & 79       & 83    & 1.01 	& 46	& 68 \\
Lat+s	    & \bf{1.80}	& \bf{27,844} & \bf{14,900}  & \bf{1.79}	& \bf{16,923}   & \bf{5,066} & \bf{1.65}	& \bf{7,290}	& \bf{2,114} \\
Lat+s+b		& 1.61	& 14,431 & 7,026   & 1.64	& 5,583    & 1,629 & 1.42	& 1,970	& 487  \\
Sob+LMS		& 1.63	& 14,812 & 7,748   & 1.62 & 8,090    & 2,328 & 1.58	& 4,197	& 1,140 \\
\hline\hline
%
 $(T,s,\tau)  \longrightarrow $ 
	 & \multicolumn{3}{c|}{$(25.6,\,128,\,0.2)$}   & \multicolumn{3}{c|}{$(102.4,\ 128,\,0.8)$} 
	 & \multicolumn{3}{c}{$(819.2,\, 1024,\, 0.8)$}  \\
\hline	
 MC $\Var$ &\multicolumn{3}{c|}{$111.0$}     &\multicolumn{3}{c|}{$166.7$} 
	 &\multicolumn{3}{c}{$166.6$}\\
\hline					
Point sets  & $ \hat\beta $ & $\vrf$ & $\er$ & $ \hat\beta $ & $\vrf$ & $\er$ & $ \hat\beta $ & $\vrf$ & $\er$  \\
\hline
MC		    & 1.00	& 1	     & 1	     & 1.00	& 1	     & 1	   & 1.00	& 1      & 1 \\
RQMC      & 1.06  & 519    & 625       & 1.10	& 2,294	 & 2,382   & 1.12	& 2,887  & 3,018	\\
Lat+s		  & \bf{1.77} & 20,206  & 11,597    & \bf{1.84}	& 34,301 & 23,364  & \bf{1.79}	& 31,160 & 22,671  \\
Lat+s+b		& 1.75 & \bf{32,136}  & \bf{16,111}    & 1.50	& 39,380 & 29,552  & 1.58	& \bf{43,977} & \bf{31,849} \\
Sob+LMS		& 1.65 & 15,709  & 8,990     & 1.66	& \bf{47,713} & \bf{33,388}  & 1.56	& 31,959 & 23,705 \\
\hline
\end{tabular} 	
\vskip 4pt
\begin{tabular}{l| crr }
\hline
 $(T,s,\tau)  \longrightarrow $  &  \multicolumn{3}{c}{$(1.6,\,8,\,0.2)$, normal} \\
\hline						
 MC $\Var$ &\multicolumn{3}{c}{$107.8$}\\
\hline
Point sets   & $ \hat\beta $ & $\vrf$ & $\eif$ \\
\hline
MC		    & 1.00	 & 1         & 1 \\
RQMC      & 1.94  & 3,673,231 & 5,484,012  \\
Lat+s	    & 1.89  & 56,510    & 8,605 \\
Lat+s+b		& 2.01 & \bf{189,471,599} & \bf{28,804,690} \\
Sob+LMS		& \bf{2.08} & 5,509,642   & 889,294 \\
\hline
\end{tabular} 	
\end{table}

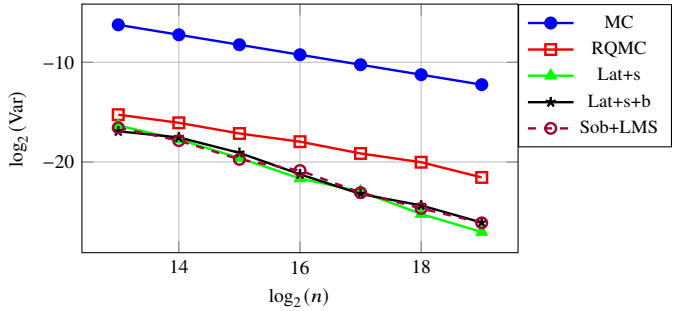
\begin{figure}[!htbp]
\begin{center} 
\begin{tikzpicture} \footnotesize 
\begin{axis}[ 
cycle list name=defaultcolorlist,
legend style={at={(1.0,1.0)},font=\scriptsize},
xlabel=$\log_2(n)$,
ylabel=$\log_2(\Var)$,
grid,
width=0.6\textwidth,
height=0.4\textwidth,
] 
      \addplot table[x=log(n),y=log(Var)] { 
      log(n)  log(Var) 
      12.999999999999998  -6.2479841662180360 
      13.999999999999998  -7.2479841662180360 
      15.0  -8.2479841662180360 
      16.0  -9.2479841662180360 
      17.0  -10.2479841662180360 
      18.0  -11.2479841662180360 
      19.0  -12.2479841662180360 
 }; 
      \addlegendentry{MC}
%
%
      \addplot table[x=log(n),y=log(Var)] { 
      log(n)  log(Var) 
      13.0	-15.25554836375491
      14.0	-16.069365279541937
      15.0	-17.14223101515822
      16.0	-17.9627114967654
      17.0	-19.142564940704776
      18.0	-20.022634819561688
      19.0	-21.54486963838362
 }; 
      \addlegendentry{RQMC}
      
      \addplot table[x=log(n),y=log(Var)] { 
      log(n)  log(Var) 
      12.999999999999998  -16.340588574087693 
      13.999999999999998  -17.741717596128925 
      15.0  -19.598164279319864 
      16.0  -21.64858120492211 
      17.0  -22.95924957640198 
      18.0  -25.20985723049743 
      19.0  -27.013037523077085 
 }; 
      \addlegendentry{Lat+s}
      \addplot table[x=log(n),y=log(Var)] { 
      log(n)  log(Var) 
      12.999999999999998  -16.919713806872945 
      13.999999999999998  -17.5378101804313 
      15.0  -19.082962211552164 
      16.0  -21.233721157458053 
      17.0  -23.204371566465287 
      18.0  -24.354716576213594 
      19.0  -26.064897060548887 
 }; 
      \addlegendentry{Lat+s+b}
      \addplot table[x=log(n),y=log(Var)] { 
      	log(n)  log(Var) 
      12.999999999999998  -16.562867951269187 
      13.999999999999998  -17.86191648024051 
      15.0  -19.736190324521953 
      16.0  -20.851232057900173 
      17.0  -23.07200697980074 
      18.0  -24.65479577168275 
      19.0  -26.102429526371164 
      }; 
      \addlegendentry{Sob+LMS}
\end{axis}
\end{tikzpicture}
\captionof{figure}{Estimated $\Var[\hat\mu_n]$ as a function of $n$,
   in log-log scale, for the reversible isomerization system, with $T=1.6$ and $s=8$.}
\label{fig:revIso}
\end{center} 
\end{figure}

Table~\ref{tab:revIso} summarizes our experimental results.
Seven cases are reported in the table.
\hflorian{The results for RQMC are now really obtained with Sobol' points. To
 have a high enough number of dimensions, I took the direction numbers from 
 \url{https://web.maths.unsw.edu.au/~fkuo/sobol/new-joe-kuo-6.21201}.}%
\hflorian{Added log-variance for MC with $n=2^{19}$. I hope this is what you wanted.}%
The first case (in the upper left) has the same parameters as \cite{sBEE19a}:
$T=1.6$, and $s=8$, so $\tau = T/s = 0.2$.
Figure~\ref{fig:revIso} illustrates how the variance decreases as a function of $n$ for this case.  
Notice the steeper slope for the four Array-RQMC variants.
Array-RQMC clearly outperforms both MC and classical RQMC in this example.

We also observe \red{from the first} three cases that when we increase $s$ 
\red{(decrease $\tau$)} with $T$ fixed, 
the factors $\vrf$ and $\eif$ diminish, and the diminution is much more prominent with RQMC.
The latter might be no surprise, because increasing $s$ increases
the dimension of the RQMC points. 
But it was unclear a priori if it would also occur with Array-RQMC, and by how much.
By doing further experimentation, we found that the decrease of \vrf{} is not
really due to the increase in the number of steps, but rather to the decrease in $\tau$.
To see this, look at the fourth case, with $(T,s,\tau) = (25.6, 128, 0.2)$.
Here we have the same $\tau$ as in the first case, but $s$ is multiplied by 16.
For the Array-RQMC methods, the variance reductions and convergence rates are similar
to the first case.
For RQMC, they are a bit lower, which is not surprising because the dimension has increased.  
\hflorian{``Almost the same'' is maybe a little bit exaggerated, but they are reasonably close.}%
For cases five and six, we have increased $\tau$ to 0.8 and we compare
two large values of $s$.  The \vrf{}'s are roughly comparable, which means that they really 
depend on $\tau$ and not much on $s$.  Why is that?

Recall that in this example, at each step we generate a pair of Poisson random variables,
which are discrete and therefore discontinuous with respect to the underlying uniforms.
The mean of each Poisson random variable is proportional to $\tau$, and the larger the mean,
the closer it is to a continuous distribution.   In fact, as $\tau$ increases, the Poisson 
converges to a normal distribution, whose inverse cdf is smooth, so the generated values 
are smooth functions of the underlying uniforms in the limit.  
That is, we obtain a better \vrf{} when $\tau$ is larger because the integrand is closer
to a continuous (and smooth) function.  When the Poisson distributions have small means, in contrast, 
the response has larger discontinuities.  And it is well known that RQMC is much more effective for 
smooth functions \red{than for} discontinuous functions.  
This kind of behavior was already pointed out for RQMC in Section 5.2 of \cite{sBEE19a}. 
Interestingly, we see that the same effect applies to Array-RQMC as well.
To illustrate this effect ``in the limit,'' we made an experiment in which all the Poisson random 
variables at each step are replaced by normals with the same mean and variance, and the state vector
has real-valued components rather than integer components, using the same parameters as in the first 
case in the table. The results are in the last (bottom) entry of the table 
and they are stunning.  Firstly, for RQMC and all Array-RQMC methods, 
the rate $\hat\beta$ is close to 2, which does not occur for the other cases.  Secondly,
the \vrf{} factor is also very large for RQMC and is huge in particular for Array-RQMC with Lat+s+b.
This surprising result for RQMC can be explained as follows. Here the integrand has
16 dimensions, but on a closer look one can see that it is a sum of 16 normal random 
variables that are almost independent;
i.e., almost a sum of one-dimensional functions.  This means that the effective dimension 
is close to 1, and this explains the success of RQMC.
\red{Essentially, only the one-dimensional projections of the points are important for 
classical RQMC, which explains the large gains for this method in this case.
For Array-RQMC, the two-dimensional projections are important, because one additional coordinate
is used for the sort, and this is why it does not beat classical RQMC for most point sets.
The huge gain obtained with Lat+s+b is an exception. It can be explained by the fact that for a smooth 
one-dimensional function, classical RQMC with Lat+s+b can provide an $\cO(n^{-4})$ convergence rate 
for the variance \citep{vHIC02a}.  
For one-dimensional smooth functions, the baker's 
transformation produces a locally antithetic effect, so it integrates exactly the piecewise
linear approximation and only higher-order error terms remain \citep{vLEC09f}.
The huge \vrf{} indicates that \red{much} of this effect carries over to Array-RQMC.}
\hflorian{Previously you had asked, whether we see $\cO(n^{-4})$ for classical RQMC with Lat+s+b. Unfortunately, no. I have tried with 3 different lattices but admittedly none of them fitted the linear variance model well and the best one had only a \vrf{} of 1.7 million (Sob+LMS gave 3.7 millions).}
\hpierre{On the other hand, the $\cO(n^{-4})$ does not carry over: we never observe this rate
 in our results!  In principle, we should observe it for classical RQMC with Lat+s+b.
 I saw that you tried classical RQMC with Sobol' points earlier, but there is no locally antithetic 
 effect with the Sobol' points.  We should observe it for classical RQMC with Lat+s+b.}
\hpierre{It would be interesting to see if this huge gain with Lat+s+b also occurs with RQMC.
 I would expect that it does.}%
\hflorian{Not compared to Sobol+LMS. For Lat+s+b we \emph{only} get a \vrf of 568,299.}%

We just saw that as a rough rule of thumb, the RQMC methods bring more gain when the 
Poisson random variables have larger means.  
We know (from Section~\ref{sec:tau}) that the mean of the Poisson random variable $D_{j,k}$
is $a_k(\bX_{j-1}) \tau$.  This mean can be increased by increasing either $\tau$ or the 
components of the state vector. 
For the present example, if we denote $\bX_{j-1} = (X_{j-1}^{(1)},X_{j-1}^{(2)})^\tr$,
the number of molecules of each of the two types at step $j-1$,
we have $a_k(\bX_{j-1}) = c_k X_{j-1}^{(k)}$ for $k=1,2$, so the Poisson
means are increased by a factor $\gamma > 1$ by either multiplying $\tau$ by $\gamma$
or multiplying the vector $\bX_{j-1}$ by $\gamma$.
We made experiments whose results agreed with that when all the components of the state
were large enough.  But if one component of $\bX_{j-1}$ is small, and we increase $\tau$
and simulate the system over a few steps, this component has a good chance of getting 
close to zero at some step, and this increases the discontinuity. 
In that situation, a larger $\tau$ can worsen the VRF.
To further test the above reasoning, we made another set of experiments 
in which the initial state $\bX_0$ had two equal 
components, exactly $X_0^{(1)} = X_0^{(2)} = (10^2 + 10^6)/2$ molecules of each type,
and we adapted the reaction rates to $c_1=c_2=100/X_0^{(1)}$, 
to keep $\EE[X_1(t)] = X_0^{(1)}$ for all $t$.
\hflorian{This keeps the reaction probabilities around the same magnitude and maintains
the relation $\EE[\bX_t]=\bx_0$ for all $t\geq0$, just as we had for the previous setting,
which is what makes this comparison \emph{fair.}}%
In this case, the problem of one component 
getting close to 0 does not occur so things remain smoother.
We found that the VRFs were larger than in Table~\ref{tab:revIso} for both RQMC and 
Array-RQMC (we exclude the normal distribution). 
The VRF for RQMC was also smaller when both $T$ and $s$ were large,
but not when $s$ was increased and $T$ remained small.
One possible explanation for this is that when $T$ and $s$ are large,
the overall change in the state can be large, and then the set of successive changes
in the state are less independent, which increases the effective dimension.

\subsection{The Schl\"{o}gl System }
\label{sec:schlogl}

\hpierre{For this example, in the end I find that the results with the original model,
with $S_2$ and $S_3$ fixed, are actually more interesting despite the fact that the chain
is unidimensional.  The results are clear, simple, and convincing.
So I propose to put them first, with the table.
After that, we can put results with our original model, which is two-dimensional.
But since the model is very similar, one expects that sorting on $S_1$ only should perform well,
and it does.  So there is no real motivation to go through complicated gymnastic to design 
more complicated sorts, just to gain a little.  I think we should keep the discussion short
in that part.}
\hpierre{The point that really tickles me in that example is something else.
 Since the two variants of the model are very similar (because the copy numbers of 
 $S_2$ and $S_3$ are very large), why is the MC variance so different between the two?
 For $T=4$ and $s=16$, the MC variance was 27,409 in our paper, but in you new results, 
 you get 107.8.   The difference is huge!   
 What about the estimated expectation?   I think it should be reported for all our examples,
 perhaps as the first row in the tables, so other people can compare their results with ours,
 and we can perhaps check if all our methods give a correct answer.
 Also, is the behavior of the trajectories (our old Fig.3) similar with the two models,
 or is it very different?}%
\hflorian{Good catch! The numbers for the MC Variance are wrong. Perhaps, I copied the 
 table from an old version and forgot to correct these values. I updated them in the 
 table now.}%

In this second example, also taken from \cite{sBEE19a}, we have the three species 
$ S_{1} $, $ S_{2} $  and $ S_{3} $,  and four reaction channels with reaction rates 
$ c_{1} = 3\times 10^{-7} $, $ c_{2} = 10^{-4} $,  $ c_{3} = 10^{-3} $  and $ c_{4} =3.5$, respectively.
The model can be depicted as:
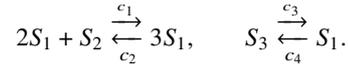
\begin{align*}
  2S_{1}+S_{2}  \stackrel{\xrightarrow{c_1}}{\xleftarrow[c_2]{}}  3S_{1},  &\qquad
  S_{3} \stackrel{\xrightarrow{c_3}}{\xleftarrow[c_4]{}} S_{1}.
\end{align*}
The propensity functions $a_k$ are given by 
\begin{align*}
  a_{1}(\bx) &= c_{1}x_1(x_1-1)x_2 /2, &
  a_{2}(\bx) &= c_{2}x_1(x_1-1)(x_1-2) /6, \\
  a_{3}(\bx) &= c_{3}x_3, &  
  a_{4}(\bx) &= c_{4}x_1.
\end{align*}
We also take $\bx_{0}= (250,\, 10^{5},\, 2\times 10^{5})^\tr$, $ T=4$, 
and $\tau = 1/4$, so $s = 16$ steps.
This is the same model as in \cite{sBEE19a}, with the same parameters, except that we took a 
slightly smaller $\tau$ to avoid negative copy numbers \red{(they had $\tau = 0.4$ also with $T=4$).
As in \cite{sBEE19a}, we also make the simplifying assumption that the copy numbers of 
$S_2$ and $S_3$ never change.  
\hflorian{Should we mention that this makes the system bistable. I suppose this is the main
motivation for this assumption, rather than that it makes the system simpler.}%
Only $X_1(t)$, the copy number of $S_1$, is changing
when reactions occur.  (We will relax this assumption later.)
Under this assumption, the Markov chain has a one-dimensional state and the sorting
is straightforward, as in the previous example.
We want to estimate $\EE[X_1(T)]$, the expected number of molecules of $S_1$ at time $T$.
Here, this expectation does depend on $T$, and we will see that $\Var[X_1(T)]$ also
depends very much on $T$. Our aim is to compare the efficiencies of MC, RQMC, and Array-RQMC. 
With $d=4$ possible reactions, the RQMC points must have 5 dimensions for Array-RQMC
and $d s = 64$ dimensions for classical RQMC.}

\red{Table~\ref{tab:schloegl-bistable} reports experimental results for this example,
first with the parameters just mentioned, then with $(T, s, \tau) = (16, 128, 1/8)$,
and finally with $g(\bX(T)) = X_1(T)$ replaced by $g(\bX(T)) = \II[X_1(T) > 300]$,
so we estimate the probability of having more than 300 molecules of $S_1$ at time $T$
instead of the expected number.
In all cases, we see that classical RQMC does not bring much gain,
\hpierre{Is this comparable to what \cite{sBEE19a} had?}%
\hflorian{They compare the RMSE (although I am not convinced that
they really considered the RMSE, not the variance) in Fig. 13. What we see in that plot,
though, is almost exactly what we observe.}%
whereas Array-RQMC brings very large variance reductions and efficiency improvements.
The gains are larger for the first set of parameters;
for classical RQMC, this comes from the smaller dimension, whereas
for Array-RQMC, this is due to the larger $\tau$ and smaller $s$
(we made additional experiments and observed that the gains were slightly better
when we increased $\tau$ for fixed $s$ or we decreased $s$ for fixed $\tau$).
}

\begin{table}[!htbp] 
 \centering
 \caption{Estimated rates $\hat\beta$, $\vrf$, and $\eif$, 
  for the Schl\"ogl system, for various choices of $(T,s,\tau)$ and for two definitions of $g$.}
\label{tab:schloegl-bistable}
\small
\begin{tabular}{l| crr| crr | crr}
\hline
 & \multicolumn{3}{c|}{$g(\bX(t)) = X_1(t)$} & \multicolumn{3}{c|}{$g(\bX(t)) = X_1(t)$} 
 & \multicolumn{3}{c}{$g(\bX(t)) = \II[X_1(t) > 300]$} \\
\hline
 $(T,s,\tau)  \longrightarrow $ 
   & \multicolumn{3}{c|}{$(4,\,16,\,1/4)$}   & \multicolumn{3}{c|}{$(16,\,128,\,1/8)$} 
	 & \multicolumn{3}{c}{$(16,\,128,\,1/8)$}   \\
\hline						
 $\EE[g(\bX_s)]$ &\multicolumn{3}{c|}{$309.0$}     &\multicolumn{3}{c|}{$318.3$}  
           &\multicolumn{3}{c}{$0.49$}  \\ 
 \hline
 MC $\Var$ &\multicolumn{3}{c|}{44,575}     &\multicolumn{3}{c|}{56,465}  
           &\multicolumn{3}{c}{0.25}  \\  
\hline
Point sets  & $ \hat\beta $ & $\vrf$ & $\eif$ & $ \hat\beta $ & $\vrf$ & $\eif$ 
            & $ \hat\beta $ & $\vrf$ & $\eif$ \\
\hline
MC		    & 1.00	& 1      & 1       & 1.00	& 1       & 1			& 1.00	& 1	& 1 \\
RQMC      & 1.10	& 9	     & 9       & 1.04   & 3       & 3			& 1.04	& 3	& 5 \\
Lat+s	    & \bf{1.64}	& 2,897  & 2,458   & 1.62	& 1,467   & 1,369	& 1.36	& 1,273	& 1,198 \\
Lat+s+b		& 1.24	& 10,147  & 9,318    & 1.09	& 3,427   & 3,341   	& 1.07	& 2,709	& 2,535 \\
Sob+LMS		& 1.56	& \bf{15,043} & \bf{14,079}  & \bf{1.70}   & \bf{6,905}   & \bf{6,681}	& \bf{1.65}	& \bf{5,730}	& \bf{5,512} \\
\hline
\end{tabular}
\end{table}

\red{For the purpose of having a higher-dimensional state, we now consider a slightly different 
version of this model, in which the copy numbers of all molecule types are assumed to vary.  
Since the total number of molecules remains constant over time, the dimension of the state
can be taken as $\ell = 2$.  We take the state as $\bX = (X^{(1)}, X^{(2)})^\tr$, and $X^{(3)}$ can be
deduced by $X^{(3)} = N_0 - X^{(1)} - X^{(2)}$ where $N_0$ is the total number of molecules.
Given that the model discussed previously can be seen as an approximation of this altered model, 
we expect $X^{(1)}$ to be the most important variable for the sort in Array-RQMC.
Thus, it appears sensible to sort by $X^{(1)}$ alone, and we will try that.
\hpierre{Indeed, this is the simplest and most obvious option to try in this case!
  We should have an entry for this sort in the Table.}%
\hflorian{I included the values for Sob+LMS in the table, for the other two I have
started the experiments. Will add them later. I also added a line for the estimated mean.}%
We will also try other sorts based on the two-dimensional state and compare.
With $d=4$ possible reactions, the RQMC points for Array-RQMC must be five-dimensional 
if we construct an importance function $h$ that maps the state to one dimension,
and must be six-dimensional otherwise.
With classical RQMC, the dimension of the RQMC points is $d s = 64$ for the first case 
and 512 for the second case.
}
%
\hpierre{\bf I still have much polishing to do.  On the other hand, we should not try
  to discuss too many details, because it distracts the reader from the essential.}

We now examine how to construct an importance function $h_j : \NN_0^2 \to \RR$ 
as discussed in Section~\ref{sec:sorting-strategies}.
\red{With the OSLAIF,} one can compute the conditional expectation \emph{exactly} by using 
(\ref{eq:econd-onestep}). \red{This gives $h(\bx)= x_1+\tau(a_1(\bx)-a_2(\bx)+a_3(\bx)-a_4(\bx))$,
which is a polynomial in $x_1$, $x_2$, $x_3$,  with coefficients that are easy to compute.}
To obtain a SDIF for a more general $j$, one possible heuristic could be to assume
the same form of polynomial (even if this is not exact)
and select the coefficients by least-squares fitting to data obtained 
from pilot runs as explained in Section~\ref{sec:sorting-strategies}.
We did this and we also tried fitting a more general bivariate polynomial 
that contains all possible monomials $x^{\varepsilon_1}y^{\varepsilon_2}$ with 
$0\leq\varepsilon_1, \varepsilon_2\leq3$, 
but this gave us no improvement over OSLAIF. 
The other SDIF approches that we tried also did no better than OSLAIF.
A plausible explanation is that the functions $h_j$ in this case are based on 
data obtained from noisy simulations (large variance and dependence on $j$).
\hflorian{Another explanation might by the two transient regimes around $T=4$.}
\hflorian{Actually, we may check that again to make sure this has not been caused by any of the 
discovered bugs...}
\hpierre{[Explain this further, perhaps with some plots like in Figure~\ref{fig:schloegl-per-run},
  .... after we get all the results.]}
For the batch sort, we kept the three coordinates in their natural order
and we used $n_1 = n_2 = \lceil n^{1/2}\rceil$.
\hpierre{Is this ordering really the best?  Is $x_1$ really the most important coordinate?}%
\hpierre{What about using $n_1=n$, which means sorting on the first coordinate only?}
\hflorian{The referee asked for 2 things here: 1) what if we simplify the OSLAIF and 2) how 
sensitive is the SDIF w.r.t. the number of pilot samples. We steered around that pretty elegantly,
but we should probably not forget to mention something in the response letters. For 1) see the last
paragraph in my notes on page 22. For 2) my experiments suggest that the bad performance of the SDIF 
is not really due to noisy observations. In fact, in the PKA example, it did not matter at all how 
many pilot samples we used. In the Schl\"ogl system, the two transient regimes might be the problem. 
See my notes, p. 23, penultimate paragraph (right before ``The Schl\"ogl system'').}

Table~\ref{tab:schloegl} summarizes our experimental results with this example.
\hflorian{The referee suggests that we boldface the best result for each sort 
and underline the globally best performance. I think it might become hard to read.
What do you think. In case you want to implement that, I can take care of it.}%
Again, Array-RQMC performs much better than RQMC, with \vrf{}'s in the thousands.
All sorting methods reported in the table perform reasonably well.
The OSLAIF is very effective for $T=4$, but somewhat less effective for $T=32$.
The Sobol' points are generally the best performers.  

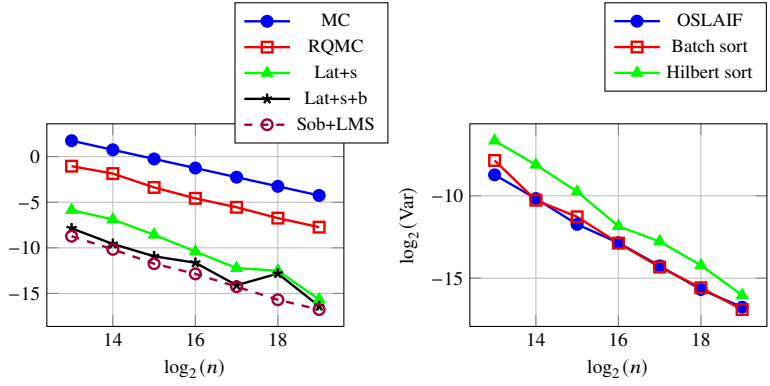
\begin{figure}[!htbp]
\scriptsize
\centering
 \begin{tikzpicture} 
    \begin{axis}[ 
      xlabel=$\log_2(n)$,
      width=0.45\columnwidth,
       height=0.35\columnwidth,
       legend style={at={(1.15,1.6)},anchor=north east},	    
       grid,
,
      ] 
      \addplot table[x=log(n),y=log(Var)] { 
      log(n)  log(Var) 
13	1.7424
14	0.742404
15	-0.257596
16	-1.2576
17	-2.2576
18	-3.2576
19	-4.2576
 }; 
      \addlegendentry{MC}
      \addplot table[x=log(n),y=log(Var)] { 
      log(n)  log(Var) 
13.0    -1.0487520726856359
14.0    -1.8616354085136353
15.0    -3.376942803446158
16.0    -4.573570379692713
17.0    -5.555605516191519
18.0    -6.740239755809068
19.0    -7.73175448974521
 }; 
      \addlegendentry{RQMC}
      \addplot table[x=log(n),y=log(Var)] { 
      log(n)  log(Var) 
      12.999999999999998  -5.858829405688029 
      13.999999999999998  -6.8954114670343145 
      15.0  -8.567087765202961 
      16.0  -10.404400823727434 
      17.0  -12.213111145730332 
      18.0  -12.50944955396411 
      19.0  -15.648684183690497 
 }; 
      \addlegendentry{Lat+s}
      \addplot table[x=log(n),y=log(Var)] { 
      log(n)  log(Var) 
      12.999999999999998  -7.862713224657488 
      13.999999999999998  -9.577251799042514 
      15.0  -10.92893904818212 
      16.0  -11.631705326707861 
      17.0  -14.09897316752091 
      18.0  -12.82080244656712 
      19.0  -16.388629172093324
 }; 
      \addlegendentry{Lat+s+b }
      \addplot table[x=log(n),y=log(Var)] { 
      log(n)  log(Var) 
      12.999999999999998  -8.727618025014852 
      13.999999999999998  -10.165915861681764 
      15.0  -11.735033278471262 
      16.0  -12.853419600166463 
      17.0  -14.257290489699082 
      18.0  -15.690023763787385 
      19.0  -16.765145197891623  
 }; 
      \addlegendentry{Sob+LMS}
%
    \end{axis}
  \end{tikzpicture}
   \begin{tikzpicture} 
    \begin{axis}[ 
	  font=\scriptsize,
      xlabel=$\log_2(n)$,
      ylabel=$\log_2(\Var)$,
       grid,
       width=0.45\columnwidth,
       height=0.35\columnwidth,
       legend style={at={(1.0,1.6)},anchor=north east},	       
      ] 
	\addplot table[x=log(n),y=log(Var)] { 
      log(n)  log(Var) 
      12.999999999999998  -8.727618025014852 
      13.999999999999998  -10.165915861681764 
      15.0  -11.735033278471262 
      16.0  -12.853419600166463 
      17.0  -14.257290489699082 
      18.0  -15.690023763787385 
      19.0  -16.765145197891623 
 }; 
      \addlegendentry{OSLAIF}
      \addplot table[x=log(n),y=log(Var)] { 
      log(n)  log(Var) 
      12.999999999999998  -7.855266652951311 
      13.999999999999998  -10.26935540021282 
      15.0  -11.275948737471586 
      16.0  -12.874794042215099 
      17.0  -14.328223008818942 
      18.0  -15.584546377734254 
      19.0  -16.909000323507232
 }; 
      \addlegendentry{Batch sort}
%
%
      \addplot table[x=log(n),y=log(Var)] { 
      log(n)  log(Var) 
      12.999999999999998  -6.65340065874333 
      13.999999999999998  -8.116227966147198 
      15.0  -9.745733245496993 
      16.0  -11.844085963644343 
      17.0  -12.774946663157467 
      18.0  -14.219327357796246 
      19.0  -16.035613598438164
 }; 
      \addlegendentry{Hilbert sort}
    \end{axis} 
  \end{tikzpicture}
  \caption{Empirical variance of the sorting methods vs $n$ in a log-log scale
	 for $T=4$ and $s=16$,
	 for the OSLAIF sort and various point sets (left) and for various sorts with Sobol+LMS (right).}
  \label{fig:schloegl-var-n}
\end{figure}

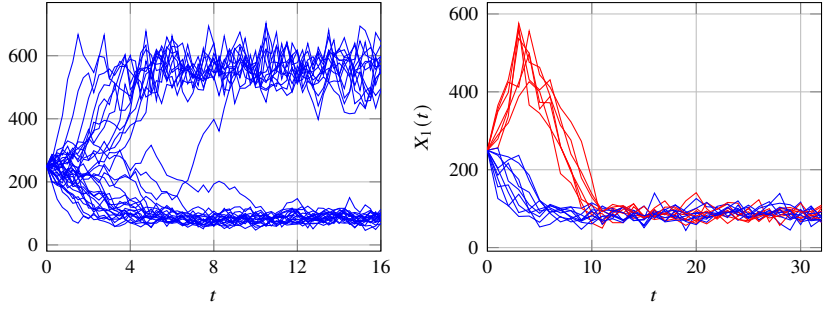
\begin{figure}
\begin{tikzpicture} 
    \begin{axis}[ 
       xlabel=$t$,
       grid,
       width=0.49\columnwidth,
       height=0.4\columnwidth,
       xmin=0,
       xmax=16,
       xtick={0,4,8,12,16},
      ] 
\addplot[blue] table[x=var1,y=var2] { 
      var1  var2 
0.	250.
0.25	221.
0.5	192.
0.75	216.
1.	221.
1.25	224.
1.5	185.
1.75	159.
2.	117.
2.25	116.
2.5	107.
2.75	111.
3.	94.
3.25	96.
3.5	92.
3.75	105.
4.	111.
4.25	104.
4.5	98.
4.75	71.
5.	55.
5.25	78.
5.5	75.
5.75	71.
6.	76.
6.25	80.
6.5	91.
6.75	95.
7.	71.
7.25	91.
7.5	96.
7.75	94.
8.	75.
8.25	74.
8.5	63.
8.75	72.
9.	78.
9.25	95.
9.5	94.
9.75	77.
10.	64.
10.25	58.
10.5	66.
10.75	75.
11.	93.
11.25	102.
11.5	109.
11.75	104.
12.	104.
12.25	88.
12.5	85.
12.75	100.
13.	83.
13.25	94.
13.5	79.
13.75	85.
14.	71.
14.25	69.
14.5	70.
14.75	56.
15.	71.
15.25	87.
15.5	85.
15.75	73.
16.	75.
};
\addplot[blue] table[x=var1,y=var2] { 
      var1  var2 
0.	250.
0.25	225.
0.5	231.
0.75	206.
1.	215.
1.25	217.
1.5	182.
1.75	184.
2.	167.
2.25	155.
2.5	142.
2.75	158.
3.	163.
3.25	123.
3.5	109.
3.75	109.
4.	86.
4.25	100.
4.5	79.
4.75	93.
5.	101.
5.25	132.
5.5	121.
5.75	116.
6.	112.
6.25	99.
6.5	105.
6.75	88.
7.	108.
7.25	88.
7.5	90.
7.75	89.
8.	88.
8.25	75.
8.5	84.
8.75	77.
9.	72.
9.25	61.
9.5	76.
9.75	70.
10.	56.
10.25	58.
10.5	64.
10.75	84.
11.	87.
11.25	87.
11.5	87.
11.75	71.
12.	68.
12.25	80.
12.5	104.
12.75	90.
13.	78.
13.25	86.
13.5	96.
13.75	96.
14.	88.
14.25	90.
14.5	81.
14.75	91.
15.	73.
15.25	46.
15.5	54.
15.75	60.
16.	82.
};
\addplot[blue] table[x=var1,y=var2] { 
      var1  var2 
0.	250.
0.25	255.
0.5	263.
0.75	277.
1.	278.
1.25	250.
1.5	246.
1.75	257.
2.	227.
2.25	205.
2.5	186.
2.75	148.
3.	128.
3.25	128.
3.5	131.
3.75	136.
4.	127.
4.25	118.
4.5	132.
4.75	104.
5.	108.
5.25	96.
5.5	116.
5.75	120.
6.	106.
6.25	99.
6.5	102.
6.75	101.
7.	94.
7.25	102.
7.5	96.
7.75	84.
8.	62.
8.25	76.
8.5	85.
8.75	87.
9.	87.
9.25	67.
9.5	78.
9.75	88.
10.	102.
10.25	91.
10.5	94.
10.75	76.
11.	89.
11.25	86.
11.5	91.
11.75	93.
12.	110.
12.25	86.
12.5	74.
12.75	51.
13.	70.
13.25	87.
13.5	79.
13.75	68.
14.	87.
14.25	77.
14.5	94.
14.75	80.
15.	96.
15.25	101.
15.5	110.
15.75	93.
16.	113.
};
\addplot[blue] table[x=var1,y=var2] { 
      var1  var2 
0.	250.
0.25	260.
0.5	225.
0.75	220.
1.	242.
1.25	241.
1.5	212.
1.75	225.
2.	221.
2.25	219.
2.5	223.
2.75	256.
3.	234.
3.25	274.
3.5	301.
3.75	258.
4.	213.
4.25	232.
4.5	186.
4.75	166.
5.	164.
5.25	143.
5.5	163.
5.75	166.
6.	158.
6.25	154.
6.5	125.
6.75	103.
7.	115.
7.25	121.
7.5	112.
7.75	93.
8.	107.
8.25	95.
8.5	96.
8.75	89.
9.	81.
9.25	88.
9.5	114.
9.75	114.
10.	96.
10.25	90.
10.5	82.
10.75	60.
11.	77.
11.25	70.
11.5	68.
11.75	90.
12.	80.
12.25	84.
12.5	72.
12.75	79.
13.	77.
13.25	84.
13.5	92.
13.75	84.
14.	93.
14.25	80.
14.5	82.
14.75	73.
15.	63.
15.25	74.
15.5	73.
15.75	63.
16.	67.
};
\addplot[blue] table[x=var1,y=var2] { 
      var1  var2 
0.	250.
0.25	242.
0.5	238.
0.75	251.
1.	218.
1.25	240.
1.5	216.
1.75	217.
2.	227.
2.25	242.
2.5	252.
2.75	249.
3.	218.
3.25	184.
3.5	179.
3.75	164.
4.	120.
4.25	118.
4.5	88.
4.75	92.
5.	83.
5.25	79.
5.5	83.
5.75	59.
6.	81.
6.25	88.
6.5	61.
6.75	77.
7.	90.
7.25	67.
7.5	65.
7.75	60.
8.	62.
8.25	66.
8.5	70.
8.75	86.
9.	104.
9.25	106.
9.5	93.
9.75	68.
10.	68.
10.25	75.
10.5	86.
10.75	87.
11.	63.
11.25	84.
11.5	89.
11.75	105.
12.	110.
12.25	90.
12.5	84.
12.75	96.
13.	82.
13.25	75.
13.5	67.
13.75	72.
14.	98.
14.25	116.
14.5	102.
14.75	90.
15.	93.
15.25	84.
15.5	92.
15.75	97.
16.	94.
};
\addplot[blue] table[x=var1,y=var2] { 
      var1  var2 
0.	250.
0.25	286.
0.5	257.
0.75	299.
1.	354.
1.25	410.
1.5	488.
1.75	530.
2.	545.
2.25	483.
2.5	539.
2.75	545.
3.	558.
3.25	538.
3.5	566.
3.75	566.
4.	601.
4.25	531.
4.5	567.
4.75	677.
5.	567.
5.25	598.
5.5	587.
5.75	659.
6.	622.
6.25	504.
6.5	542.
6.75	572.
7.	552.
7.25	506.
7.5	593.
7.75	598.
8.	577.
8.25	586.
8.5	526.
8.75	546.
9.	557.
9.25	590.
9.5	655.
9.75	569.
10.	578.
10.25	504.
10.5	494.
10.75	509.
11.	503.
11.25	591.
11.5	532.
11.75	563.
12.	516.
12.25	575.
12.5	639.
12.75	570.
13.	607.
13.25	539.
13.5	618.
13.75	597.
14.	620.
14.25	657.
14.5	601.
14.75	553.
15.	542.
15.25	492.
15.5	466.
15.75	528.
16.	547.
};
\addplot[blue] table[x=var1,y=var2] { 
      var1  var2 
0.	250.
0.25	269.
0.5	247.
0.75	268.
1.	260.
1.25	294.
1.5	306.
1.75	321.
2.	333.
2.25	363.
2.5	369.
2.75	424.
3.	466.
3.25	493.
3.5	544.
3.75	602.
4.	493.
4.25	513.
4.5	609.
4.75	633.
5.	482.
5.25	605.
5.5	621.
5.75	564.
6.	602.
6.25	583.
6.5	486.
6.75	454.
7.	482.
7.25	545.
7.5	624.
7.75	547.
8.	607.
8.25	568.
8.5	560.
8.75	565.
9.	565.
9.25	593.
9.5	575.
9.75	524.
10.	531.
10.25	540.
10.5	694.
10.75	641.
11.	553.
11.25	541.
11.5	546.
11.75	542.
12.	511.
12.25	555.
12.5	587.
12.75	597.
13.	566.
13.25	595.
13.5	659.
13.75	599.
14.	587.
14.25	582.
14.5	596.
14.75	535.
15.	578.
15.25	456.
15.5	458.
15.75	528.
16.	547.
};
\addplot[blue] table[x=var1,y=var2] { 
      var1  var2 
0.	250.
0.25	231.
0.5	244.
0.75	215.
1.	181.
1.25	192.
1.5	211.
1.75	194.
2.	171.
2.25	154.
2.5	164.
2.75	163.
3.	112.
3.25	116.
3.5	120.
3.75	131.
4.	144.
4.25	150.
4.5	133.
4.75	88.
5.	92.
5.25	87.
5.5	81.
5.75	83.
6.	93.
6.25	90.
6.5	90.
6.75	88.
7.	90.
7.25	57.
7.5	65.
7.75	63.
8.	65.
8.25	73.
8.5	80.
8.75	70.
9.	77.
9.25	88.
9.5	81.
9.75	84.
10.	82.
10.25	80.
10.5	97.
10.75	85.
11.	95.
11.25	57.
11.5	74.
11.75	82.
12.	74.
12.25	71.
12.5	82.
12.75	94.
13.	99.
13.25	101.
13.5	104.
13.75	100.
14.	89.
14.25	81.
14.5	83.
14.75	78.
15.	72.
15.25	82.
15.5	74.
15.75	80.
16.	97.
};
\addplot[blue] table[x=var1,y=var2] { 
      var1  var2 
0.	250.
0.25	280.
0.5	248.
0.75	219.
1.	234.
1.25	292.
1.5	264.
1.75	281.
2.	339.
2.25	318.
2.5	311.
2.75	292.
3.	273.
3.25	284.
3.5	281.
3.75	267.
4.	236.
4.25	258.
4.5	259.
4.75	296.
5.	316.
5.25	281.
5.5	253.
5.75	221.
6.	214.
6.25	207.
6.5	204.
6.75	194.
7.	181.
7.25	168.
7.5	190.
7.75	167.
8.	193.
8.25	202.
8.5	194.
8.75	176.
9.	184.
9.25	163.
9.5	153.
9.75	140.
10.	118.
10.25	122.
10.5	101.
10.75	106.
11.	96.
11.25	107.
11.5	133.
11.75	136.
12.	147.
12.25	122.
12.5	101.
12.75	108.
13.	100.
13.25	95.
13.5	101.
13.75	111.
14.	115.
14.25	98.
14.5	97.
14.75	84.
15.	91.
15.25	89.
15.5	106.
15.75	100.
16.	101.
};
\addplot[blue] table[x=var1,y=var2] { 
      var1  var2 
0.	250.
0.25	244.
0.5	285.
0.75	254.
1.	232.
1.25	256.
1.5	227.
1.75	224.
2.	230.
2.25	213.
2.5	198.
2.75	229.
3.	220.
3.25	206.
3.5	211.
3.75	217.
4.	211.
4.25	243.
4.5	221.
4.75	231.
5.	211.
5.25	199.
5.5	169.
5.75	169.
6.	143.
6.25	157.
6.5	193.
6.75	204.
7.	255.
7.25	314.
7.5	352.
7.75	374.
8.	398.
8.25	361.
8.5	434.
8.75	494.
9.	526.
9.25	578.
9.5	578.
9.75	555.
10.	572.
10.25	630.
10.5	577.
10.75	520.
11.	563.
11.25	612.
11.5	633.
11.75	572.
12.	578.
12.25	618.
12.5	542.
12.75	582.
13.	613.
13.25	581.
13.5	681.
13.75	577.
14.	535.
14.25	597.
14.5	606.
14.75	632.
15.	584.
15.25	555.
15.5	566.
15.75	673.
16.	644.
};
\addplot[blue] table[x=var1,y=var2] { 
      var1  var2 
0.	250.
0.25	243.
0.5	301.
0.75	308.
1.	330.
1.25	322.
1.5	305.
1.75	301.
2.	354.
2.25	424.
2.5	480.
2.75	508.
3.	505.
3.25	548.
3.5	582.
3.75	568.
4.	549.
4.25	529.
4.5	488.
4.75	542.
5.	555.
5.25	458.
5.5	563.
5.75	577.
6.	630.
6.25	627.
6.5	640.
6.75	578.
7.	532.
7.25	536.
7.5	585.
7.75	694.
8.	638.
8.25	535.
8.5	581.
8.75	621.
9.	591.
9.25	530.
9.5	589.
9.75	568.
10.	509.
10.25	530.
10.5	515.
10.75	550.
11.	556.
11.25	579.
11.5	641.
11.75	580.
12.	615.
12.25	544.
12.5	550.
12.75	549.
13.	564.
13.25	540.
13.5	604.
13.75	650.
14.	540.
14.25	580.
14.5	542.
14.75	448.
15.	462.
15.25	487.
15.5	500.
15.75	474.
16.	496.
};
\addplot[blue] table[x=var1,y=var2] { 
      var1  var2 
0.	250.
0.25	287.
0.5	282.
0.75	290.
1.	285.
1.25	307.
1.5	283.
1.75	275.
2.	282.
2.25	292.
2.5	257.
2.75	299.
3.	325.
3.25	358.
3.5	395.
3.75	509.
4.	457.
4.25	509.
4.5	556.
4.75	550.
5.	580.
5.25	573.
5.5	576.
5.75	530.
6.	534.
6.25	475.
6.5	522.
6.75	514.
7.	517.
7.25	493.
7.5	491.
7.75	561.
8.	546.
8.25	590.
8.5	493.
8.75	490.
9.	620.
9.25	636.
9.5	606.
9.75	557.
10.	499.
10.25	502.
10.5	616.
10.75	606.
11.	674.
11.25	546.
11.5	535.
11.75	485.
12.	531.
12.25	570.
12.5	493.
12.75	451.
13.	398.
13.25	478.
13.5	517.
13.75	489.
14.	474.
14.25	421.
14.5	416.
14.75	418.
15.	452.
15.25	433.
15.5	491.
15.75	592.
16.	562.
};
\addplot[blue] table[x=var1,y=var2] { 
      var1  var2 
0.	250.
0.25	234.
0.5	219.
0.75	183.
1.	201.
1.25	194.
1.5	194.
1.75	215.
2.	234.
2.25	251.
2.5	266.
2.75	263.
3.	293.
3.25	292.
3.5	297.
3.75	311.
4.	340.
4.25	392.
4.5	394.
4.75	420.
5.	560.
5.25	491.
5.5	449.
5.75	509.
6.	504.
6.25	532.
6.5	549.
6.75	564.
7.	498.
7.25	573.
7.5	528.
7.75	529.
8.	576.
8.25	564.
8.5	476.
8.75	560.
9.	560.
9.25	500.
9.5	512.
9.75	602.
10.	532.
10.25	550.
10.5	519.
10.75	530.
11.	488.
11.25	498.
11.5	474.
11.75	486.
12.	568.
12.25	533.
12.5	526.
12.75	539.
13.	519.
13.25	558.
13.5	604.
13.75	610.
14.	585.
14.25	588.
14.5	582.
14.75	490.
15.	477.
15.25	435.
15.5	497.
15.75	434.
16.	490.
};
\addplot[blue] table[x=var1,y=var2] { 
      var1  var2 
0.	250.
0.25	267.
0.5	282.
0.75	248.
1.	293.
1.25	255.
1.5	242.
1.75	208.
2.	209.
2.25	226.
2.5	219.
2.75	213.
3.	186.
3.25	191.
3.5	214.
3.75	223.
4.	209.
4.25	200.
4.5	159.
4.75	136.
5.	84.
5.25	87.
5.5	84.
5.75	93.
6.	77.
6.25	77.
6.5	82.
6.75	77.
7.	83.
7.25	86.
7.5	75.
7.75	79.
8.	59.
8.25	87.
8.5	89.
8.75	76.
9.	90.
9.25	79.
9.5	95.
9.75	100.
10.	92.
10.25	104.
10.5	82.
10.75	81.
11.	97.
11.25	95.
11.5	79.
11.75	75.
12.	69.
12.25	86.
12.5	81.
12.75	84.
13.	87.
13.25	101.
13.5	96.
13.75	83.
14.	103.
14.25	108.
14.5	90.
14.75	100.
15.	92.
15.25	85.
15.5	106.
15.75	100.
16.	83.
};
\addplot[blue] table[x=var1,y=var2] { 
      var1  var2 
0.	250.
0.25	254.
0.5	263.
0.75	267.
1.	281.
1.25	300.
1.5	327.
1.75	325.
2.	366.
2.25	323.
2.5	350.
2.75	373.
3.	416.
3.25	423.
3.5	514.
3.75	570.
4.	589.
4.25	583.
4.5	463.
4.75	501.
5.	482.
5.25	483.
5.5	597.
5.75	552.
6.	498.
6.25	530.
6.5	567.
6.75	635.
7.	563.
7.25	532.
7.5	519.
7.75	565.
8.	592.
8.25	539.
8.5	544.
8.75	547.
9.	545.
9.25	628.
9.5	566.
9.75	559.
10.	615.
10.25	497.
10.5	490.
10.75	503.
11.	513.
11.25	508.
11.5	522.
11.75	640.
12.	607.
12.25	558.
12.5	555.
12.75	523.
13.	536.
13.25	567.
13.5	520.
13.75	601.
14.	623.
14.25	594.
14.5	593.
14.75	549.
15.	490.
15.25	575.
15.5	584.
15.75	557.
16.	584.
};
\addplot[blue] table[x=var1,y=var2] { 
      var1  var2 
0.	250.
0.25	250.
0.5	249.
0.75	259.
1.	257.
1.25	286.
1.5	291.
1.75	317.
2.	336.
2.25	358.
2.5	440.
2.75	463.
3.	445.
3.25	511.
3.5	427.
3.75	427.
4.	450.
4.25	534.
4.5	517.
4.75	469.
5.	479.
5.25	521.
5.5	516.
5.75	622.
6.	652.
6.25	634.
6.5	510.
6.75	558.
7.	528.
7.25	555.
7.5	615.
7.75	598.
8.	546.
8.25	551.
8.5	548.
8.75	514.
9.	546.
9.25	537.
9.5	599.
9.75	510.
10.	560.
10.25	491.
10.5	453.
10.75	533.
11.	547.
11.25	638.
11.5	565.
11.75	602.
12.	513.
12.25	561.
12.5	570.
12.75	538.
13.	626.
13.25	587.
13.5	591.
13.75	560.
14.	577.
14.25	559.
14.5	534.
14.75	583.
15.	647.
15.25	603.
15.5	593.
15.75	583.
16.	598.
};
\addplot[blue] table[x=var1,y=var2] { 
      var1  var2 
0.	250.
0.25	253.
0.5	243.
0.75	246.
1.	189.
1.25	175.
1.5	169.
1.75	160.
2.	172.
2.25	167.
2.5	170.
2.75	167.
3.	143.
3.25	148.
3.5	149.
3.75	159.
4.	142.
4.25	135.
4.5	116.
4.75	101.
5.	114.
5.25	95.
5.5	86.
5.75	86.
6.	86.
6.25	75.
6.5	68.
6.75	56.
7.	65.
7.25	77.
7.5	64.
7.75	72.
8.	60.
8.25	79.
8.5	70.
8.75	75.
9.	76.
9.25	91.
9.5	84.
9.75	85.
10.	73.
10.25	77.
10.5	84.
10.75	82.
11.	89.
11.25	90.
11.5	90.
11.75	95.
12.	88.
12.25	92.
12.5	73.
12.75	83.
13.	95.
13.25	95.
13.5	103.
13.75	98.
14.	95.
14.25	89.
14.5	100.
14.75	103.
15.	99.
15.25	81.
15.5	70.
15.75	85.
16.	105.
};
\addplot[blue] table[x=var1,y=var2] { 
      var1  var2 
0.	250.
0.25	228.
0.5	214.
0.75	168.
1.	154.
1.25	132.
1.5	136.
1.75	104.
2.	96.
2.25	80.
2.5	82.
2.75	88.
3.	80.
3.25	82.
3.5	82.
3.75	100.
4.	99.
4.25	96.
4.5	110.
4.75	120.
5.	118.
5.25	116.
5.5	106.
5.75	74.
6.	74.
6.25	74.
6.5	68.
6.75	70.
7.	67.
7.25	68.
7.5	57.
7.75	60.
8.	79.
8.25	66.
8.5	52.
8.75	61.
9.	64.
9.25	62.
9.5	72.
9.75	89.
10.	103.
10.25	110.
10.5	91.
10.75	103.
11.	84.
11.25	100.
11.5	106.
11.75	117.
12.	96.
12.25	86.
12.5	83.
12.75	80.
13.	76.
13.25	70.
13.5	86.
13.75	82.
14.	68.
14.25	85.
14.5	65.
14.75	68.
15.	87.
15.25	85.
15.5	91.
15.75	73.
16.	79.
};
\addplot[blue] table[x=var1,y=var2] { 
      var1  var2 
0.	250.
0.25	202.
0.5	189.
0.75	182.
1.	170.
1.25	161.
1.5	143.
1.75	149.
2.	135.
2.25	127.
2.5	90.
2.75	94.
3.	79.
3.25	87.
3.5	104.
3.75	110.
4.	96.
4.25	88.
4.5	73.
4.75	67.
5.	85.
5.25	110.
5.5	96.
5.75	70.
6.	69.
6.25	71.
6.5	61.
6.75	66.
7.	72.
7.25	81.
7.5	80.
7.75	82.
8.	80.
8.25	93.
8.5	116.
8.75	95.
9.	79.
9.25	84.
9.5	96.
9.75	86.
10.	67.
10.25	71.
10.5	82.
10.75	75.
11.	73.
11.25	75.
11.5	101.
11.75	114.
12.	113.
12.25	122.
12.5	88.
12.75	80.
13.	81.
13.25	72.
13.5	88.
13.75	86.
14.	90.
14.25	103.
14.5	68.
14.75	74.
15.	78.
15.25	75.
15.5	89.
15.75	89.
16.	93.
};
\addplot[blue] table[x=var1,y=var2] { 
      var1  var2 
0.	250.
0.25	278.
0.5	257.
0.75	254.
1.	274.
1.25	238.
1.5	227.
1.75	306.
2.	269.
2.25	299.
2.5	290.
2.75	293.
3.	304.
3.25	309.
3.5	332.
3.75	379.
4.	415.
4.25	432.
4.5	482.
4.75	583.
5.	526.
5.25	582.
5.5	631.
5.75	600.
6.	547.
6.25	560.
6.5	552.
6.75	553.
7.	575.
7.25	552.
7.5	499.
7.75	545.
8.	529.
8.25	571.
8.5	662.
8.75	583.
9.	576.
9.25	588.
9.5	532.
9.75	523.
10.	514.
10.25	544.
10.5	549.
10.75	580.
11.	579.
11.25	571.
11.5	550.
11.75	512.
12.	550.
12.25	596.
12.5	614.
12.75	557.
13.	595.
13.25	572.
13.5	614.
13.75	567.
14.	498.
14.25	448.
14.5	484.
14.75	554.
15.	560.
15.25	582.
15.5	582.
15.75	508.
16.	560.
};
\addplot[blue] table[x=var1,y=var2] { 
      var1  var2 
0.	250.
0.25	269.
0.5	268.
0.75	303.
1.	287.
1.25	285.
1.5	271.
1.75	302.
2.	327.
2.25	389.
2.5	383.
2.75	395.
3.	392.
3.25	382.
3.5	364.
3.75	381.
4.	424.
4.25	531.
4.5	553.
4.75	563.
5.	558.
5.25	613.
5.5	581.
5.75	628.
6.	607.
6.25	615.
6.5	545.
6.75	608.
7.	560.
7.25	597.
7.5	545.
7.75	561.
8.	524.
8.25	598.
8.5	555.
8.75	573.
9.	623.
9.25	601.
9.5	557.
9.75	544.
10.	542.
10.25	504.
10.5	539.
10.75	545.
11.	565.
11.25	577.
11.5	581.
11.75	601.
12.	618.
12.25	499.
12.5	539.
12.75	496.
13.	473.
13.25	498.
13.5	554.
13.75	584.
14.	539.
14.25	496.
14.5	565.
14.75	506.
15.	549.
15.25	511.
15.5	478.
15.75	547.
16.	599.
};
\addplot[blue] table[x=var1,y=var2] { 
      var1  var2 
0.	250.
0.25	224.
0.5	193.
0.75	189.
1.	171.
1.25	147.
1.5	153.
1.75	131.
2.	125.
2.25	111.
2.5	115.
2.75	115.
3.	109.
3.25	95.
3.5	81.
3.75	101.
4.	88.
4.25	74.
4.5	88.
4.75	87.
5.	84.
5.25	103.
5.5	109.
5.75	125.
6.	111.
6.25	97.
6.5	91.
6.75	71.
7.	69.
7.25	64.
7.5	71.
7.75	83.
8.	83.
8.25	99.
8.5	111.
8.75	114.
9.	84.
9.25	93.
9.5	90.
9.75	65.
10.	47.
10.25	58.
10.5	66.
10.75	76.
11.	60.
11.25	70.
11.5	63.
11.75	65.
12.	66.
12.25	68.
12.5	66.
12.75	69.
13.	68.
13.25	77.
13.5	63.
13.75	72.
14.	98.
14.25	88.
14.5	76.
14.75	80.
15.	69.
15.25	89.
15.5	103.
15.75	91.
16.	79.
};
\addplot[blue] table[x=var1,y=var2] { 
      var1  var2 
0.	250.
0.25	236.
0.5	209.
0.75	197.
1.	240.
1.25	226.
1.5	219.
1.75	163.
2.	165.
2.25	112.
2.5	123.
2.75	145.
3.	120.
3.25	109.
3.5	113.
3.75	82.
4.	70.
4.25	65.
4.5	57.
4.75	68.
5.	57.
5.25	79.
5.5	81.
5.75	70.
6.	72.
6.25	82.
6.5	86.
6.75	81.
7.	89.
7.25	91.
7.5	76.
7.75	84.
8.	98.
8.25	90.
8.5	91.
8.75	77.
9.	64.
9.25	87.
9.5	86.
9.75	74.
10.	72.
10.25	74.
10.5	59.
10.75	81.
11.	90.
11.25	96.
11.5	88.
11.75	111.
12.	112.
12.25	112.
12.5	123.
12.75	110.
13.	97.
13.25	107.
13.5	101.
13.75	106.
14.	85.
14.25	71.
14.5	77.
14.75	66.
15.	78.
15.25	77.
15.5	74.
15.75	71.
16.	95.
};
\addplot[blue] table[x=var1,y=var2] { 
      var1  var2 
0.	250.
0.25	255.
0.5	243.
0.75	254.
1.	231.
1.25	193.
1.5	180.
1.75	149.
2.	129.
2.25	104.
2.5	119.
2.75	110.
3.	108.
3.25	99.
3.5	94.
3.75	116.
4.	111.
4.25	109.
4.5	114.
4.75	94.
5.	90.
5.25	90.
5.5	105.
5.75	93.
6.	86.
6.25	79.
6.5	94.
6.75	73.
7.	83.
7.25	92.
7.5	98.
7.75	117.
8.	100.
8.25	97.
8.5	108.
8.75	110.
9.	120.
9.25	121.
9.5	106.
9.75	120.
10.	80.
10.25	83.
10.5	63.
10.75	79.
11.	102.
11.25	111.
11.5	114.
11.75	115.
12.	91.
12.25	101.
12.5	98.
12.75	91.
13.	117.
13.25	116.
13.5	106.
13.75	93.
14.	109.
14.25	108.
14.5	108.
14.75	88.
15.	90.
15.25	98.
15.5	91.
15.75	84.
16.	89.
};
\addplot[blue] table[x=var1,y=var2] { 
      var1  var2 
0.	250.
0.25	199.
0.5	197.
0.75	183.
1.	166.
1.25	158.
1.5	119.
1.75	93.
2.	85.
2.25	96.
2.5	86.
2.75	82.
3.	91.
3.25	71.
3.5	83.
3.75	87.
4.	102.
4.25	89.
4.5	87.
4.75	66.
5.	68.
5.25	74.
5.5	91.
5.75	82.
6.	117.
6.25	91.
6.5	88.
6.75	77.
7.	64.
7.25	61.
7.5	51.
7.75	89.
8.	90.
8.25	62.
8.5	60.
8.75	64.
9.	73.
9.25	76.
9.5	67.
9.75	85.
10.	105.
10.25	83.
10.5	85.
10.75	83.
11.	86.
11.25	92.
11.5	86.
11.75	74.
12.	69.
12.25	74.
12.5	79.
12.75	75.
13.	92.
13.25	82.
13.5	90.
13.75	105.
14.	98.
14.25	103.
14.5	95.
14.75	89.
15.	82.
15.25	85.
15.5	91.
15.75	78.
16.	82.
};
\addplot[blue] table[x=var1,y=var2] { 
      var1  var2 
0.	250.
0.25	273.
0.5	239.
0.75	228.
1.	222.
1.25	185.
1.5	189.
1.75	176.
2.	156.
2.25	146.
2.5	129.
2.75	110.
3.	89.
3.25	58.
3.5	67.
3.75	76.
4.	99.
4.25	90.
4.5	81.
4.75	101.
5.	98.
5.25	94.
5.5	87.
5.75	75.
6.	69.
6.25	80.
6.5	65.
6.75	63.
7.	77.
7.25	71.
7.5	63.
7.75	74.
8.	71.
8.25	77.
8.5	87.
8.75	88.
9.	60.
9.25	85.
9.5	100.
9.75	89.
10.	91.
10.25	92.
10.5	101.
10.75	77.
11.	74.
11.25	78.
11.5	67.
11.75	80.
12.	81.
12.25	76.
12.5	54.
12.75	83.
13.	89.
13.25	108.
13.5	116.
13.75	109.
14.	116.
14.25	116.
14.5	101.
14.75	70.
15.	62.
15.25	62.
15.5	87.
15.75	97.
16.	98.
};
\addplot[blue] table[x=var1,y=var2] { 
      var1  var2 
0.	250.
0.25	249.
0.5	261.
0.75	211.
1.	229.
1.25	238.
1.5	230.
1.75	208.
2.	172.
2.25	156.
2.5	166.
2.75	150.
3.	123.
3.25	110.
3.5	102.
3.75	80.
4.	70.
4.25	73.
4.5	78.
4.75	89.
5.	81.
5.25	96.
5.5	95.
5.75	96.
6.	105.
6.25	108.
6.5	102.
6.75	101.
7.	88.
7.25	71.
7.5	83.
7.75	65.
8.	72.
8.25	83.
8.5	64.
8.75	52.
9.	62.
9.25	62.
9.5	52.
9.75	60.
10.	66.
10.25	93.
10.5	93.
10.75	72.
11.	68.
11.25	73.
11.5	72.
11.75	68.
12.	73.
12.25	60.
12.5	70.
12.75	67.
13.	69.
13.25	62.
13.5	70.
13.75	87.
14.	89.
14.25	110.
14.5	150.
14.75	118.
15.	103.
15.25	127.
15.5	115.
15.75	82.
16.	97.
};
\addplot[blue] table[x=var1,y=var2] { 
      var1  var2 
0.	250.
0.25	273.
0.5	292.
0.75	285.
1.	313.
1.25	274.
1.5	335.
1.75	396.
2.	465.
2.25	511.
2.5	595.
2.75	665.
3.	528.
3.25	564.
3.5	571.
3.75	602.
4.	511.
4.25	550.
4.5	560.
4.75	561.
5.	572.
5.25	620.
5.5	592.
5.75	549.
6.	663.
6.25	595.
6.5	567.
6.75	554.
7.	537.
7.25	521.
7.5	531.
7.75	564.
8.	589.
8.25	551.
8.5	568.
8.75	619.
9.	538.
9.25	550.
9.5	575.
9.75	545.
10.	571.
10.25	615.
10.5	656.
10.75	545.
11.	631.
11.25	596.
11.5	643.
11.75	558.
12.	544.
12.25	583.
12.5	521.
12.75	508.
13.	441.
13.25	487.
13.5	544.
13.75	602.
14.	613.
14.25	619.
14.5	585.
14.75	589.
15.	694.
15.25	680.
15.5	636.
15.75	565.
16.	569.
};
\addplot[blue] table[x=var1,y=var2] { 
      var1  var2 
0.	250.
0.25	269.
0.5	341.
0.75	410.
1.	525.
1.25	581.
1.5	667.
1.75	603.
2.	575.
2.25	547.
2.5	599.
2.75	595.
3.	528.
3.25	513.
3.5	496.
3.75	485.
4.	472.
4.25	537.
4.5	588.
4.75	620.
5.	603.
5.25	553.
5.5	558.
5.75	520.
6.	555.
6.25	567.
6.5	588.
6.75	511.
7.	543.
7.25	527.
7.5	522.
7.75	531.
8.	533.
8.25	474.
8.5	535.
8.75	605.
9.	637.
9.25	675.
9.5	609.
9.75	571.
10.	507.
10.25	590.
10.5	621.
10.75	537.
11.	524.
11.25	501.
11.5	494.
11.75	558.
12.	519.
12.25	524.
12.5	578.
12.75	507.
13.	514.
13.25	533.
13.5	515.
13.75	520.
14.	518.
14.25	567.
14.5	593.
14.75	640.
15.	568.
15.25	643.
15.5	555.
15.75	597.
16.	543.
};
\addplot[blue] table[x=var1,y=var2] { 
      var1  var2 
0.	250.
0.25	197.
0.5	145.
0.75	119.
1.	106.
1.25	77.
1.5	69.
1.75	91.
2.	83.
2.25	101.
2.5	79.
2.75	75.
3.	69.
3.25	59.
3.5	60.
3.75	67.
4.	81.
4.25	68.
4.5	70.
4.75	72.
5.	69.
5.25	69.
5.5	80.
5.75	83.
6.	81.
6.25	86.
6.5	100.
6.75	90.
7.	93.
7.25	93.
7.5	75.
7.75	89.
8.	74.
8.25	82.
8.5	79.
8.75	97.
9.	82.
9.25	72.
9.5	86.
9.75	80.
10.	69.
10.25	70.
10.5	61.
10.75	62.
11.	59.
11.25	75.
11.5	70.
11.75	81.
12.	94.
12.25	79.
12.5	90.
12.75	91.
13.	99.
13.25	70.
13.5	80.
13.75	75.
14.	93.
14.25	84.
14.5	70.
14.75	54.
15.	72.
15.25	60.
15.5	62.
15.75	49.
16.	69.
};
\addplot[blue] table[x=var1,y=var2] { 
      var1  var2 
0.	250.
0.25	258.
0.5	247.
0.75	219.
1.	204.
1.25	238.
1.5	222.
1.75	220.
2.	223.
2.25	217.
2.5	206.
2.75	170.
3.	132.
3.25	115.
3.5	138.
3.75	128.
4.	126.
4.25	133.
4.5	138.
4.75	154.
5.	142.
5.25	126.
5.5	101.
5.75	109.
6.	90.
6.25	80.
6.5	93.
6.75	73.
7.	79.
7.25	69.
7.5	80.
7.75	71.
8.	95.
8.25	100.
8.5	87.
8.75	103.
9.	98.
9.25	87.
9.5	82.
9.75	71.
10.	87.
10.25	82.
10.5	95.
10.75	87.
11.	88.
11.25	68.
11.5	78.
11.75	93.
12.	107.
12.25	89.
12.5	83.
12.75	86.
13.	85.
13.25	79.
13.5	82.
13.75	91.
14.	95.
14.25	100.
14.5	93.
14.75	91.
15.	73.
15.25	76.
15.5	82.
15.75	105.
16.	75.
};
\addplot[blue] table[x=var1,y=var2] { 
      var1  var2 
0.	250.
0.25	230.
0.5	217.
0.75	233.
1.	215.
1.25	198.
1.5	242.
1.75	272.
2.	268.
2.25	281.
2.5	299.
2.75	328.
3.	300.
3.25	334.
3.5	356.
3.75	355.
4.	379.
4.25	389.
4.5	453.
4.75	494.
5.	565.
5.25	608.
5.5	598.
5.75	520.
6.	579.
6.25	610.
6.5	559.
6.75	568.
7.	539.
7.25	528.
7.5	544.
7.75	601.
8.	640.
8.25	547.
8.5	588.
8.75	557.
9.	449.
9.25	497.
9.5	497.
9.75	509.
10.	532.
10.25	612.
10.5	703.
10.75	584.
11.	631.
11.25	526.
11.5	577.
11.75	585.
12.	546.
12.25	484.
12.5	519.
12.75	504.
13.	539.
13.25	550.
13.5	560.
13.75	546.
14.	540.
14.25	466.
14.5	486.
14.75	565.
15.	591.
15.25	611.
15.5	626.
15.75	601.
16.	634.
};
%
%
\end{axis}
\end{tikzpicture}
%
 \begin{tikzpicture} 
    \begin{axis}[ 
       xlabel=$t$,
       ylabel=$X_1(t)$,
       grid,
        width=0.49\columnwidth,
       height=0.4\columnwidth,
       xmin=0,
       xmax=32,
,
      ] 
 	\addplot[blue] table[x=var1,y=var2] { 
      var1  var2 
0.	250.
1.	126.
2.	108.
3.	80.
4.	81.
5.	65.
6.	91.
7.	100.
8.	95.
9.	84.
10.	69.
11.	80.
12.	75.
13.	106.
14.	95.
15.	74.
16.	86.
17.	92.
18.	75.
19.	101.
20.	126.
21.	105.
22.	103.
23.	81.
24.	91.
25.	77.
26.	91.
27.	81.
28.	91.
29.	98.
30.	66.
31.	72.
32	79 
 }; 
      \addplot[red]  table[x=var1,y=var2] { 
      var1  var2 
0.	250.
1.	289.
2.	320.
3.	388.
4.	426.
5.	401.
6.	313.
7.	190.
8.	166.
9.	116.
10.	63.
11.	50.
12.	102.
13.	89.
14.	83.
15.	82.
16.	72.
17.	77.
18.	80.
19.	74.
20.	95.
21.	75.
22.	86.
23.	62.
24.	81.
25.	100.
26.	70.
27.	85.
28.	79.
29.	87.
30.	67.
31.	83.
32	65.
 }; 
      \addplot[blue]  table[x=var1,y=var2] { 
      var1  var2 
0.	250.
1.	204.
2.	225.
3.	237.
4.	172.
5.	120.
6.	91.
7.	95.
8.	104.
9.	101.
10.	115.
11.	101.
12.	101.
13.	68.
14.	96.
15.	81.
16.	81.
17.	68.
18.	95.
19.	44.
20.	89.
21.	89.
22.	84.
23.	90.
24.	96.
25.	104.
26.	71.
27.	89.
28.	62.
29.	45.
30.	91.
31.	100.
32	65
 }; 
      \addplot[blue]  table[x=var1,y=var2] { 
      var1  var2 
0.	250.
1.	177.
2.	107.
3.	137.
4.	98.
5.	53.
6.	60.
7.	80.
8.	66.
9.	72.
10.	82.
11.	107.
12.	92.
13.	104.
14.	70.
15.	71.
16.	95.
17.	84.
18.	109.
19.	93.
20.	74.
21.	65.
22.	89.
23.	94.
24.	70.
25.	88.
26.	87.
27.	92.
28.	90.
29.	93.
30.	117.
31.	54.
32	73
 }; 
      \addplot[red]  table[x=var1,y=var2] { 
      var1  var2 
0.	250.
1.	256.
2.	337.
3.	568.
4.	366.
5.	345.
6.	307.
7.	259.
8.	201.
9.	101.
10.	85.
11.	66.
12.	97.
13.	88.
14.	113.
15.	75.
16.	102.
17.	120.
18.	88.
19.	84.
20.	67.
21.	116.
22.	98.
23.	112.
24.	81.
25.	101.
26.	123.
27.	66.
28.	98.
29.	75.
30.	78.
31.	85.
32	71
 }; 
      \addplot[red]  table[x=var1,y=var2] { 
      var1  var2 
0.	250.
1.	303.
2.	359.
3.	537.
4.	499.
5.	373.
6.	372.
7.	260.
8.	195.
9.	144.
10.	90.
11.	87.
12.	64.
13.	64.
14.	105.
15.	82.
16.	91.
17.	88.
18.	100.
19.	85.
20.	101.
21.	79.
22.	84.
23.	80.
24.	73.
25.	65.
26.	89.
27.	100.
28.	57.
29.	88.
30.	96.
31.	72.
32	93 
 }; 
      \addplot[red]  table[x=var1,y=var2] { 
      var1  var2 
0.	250.
1.	355.
2.	425.
3.	412.
4.	556.
5.	444.
6.	412.
7.	359.
8.	323.
9.	273.
10.	176.
11.	60.
12.	88.
13.	105.
14.	75.
15.	82.
16.	83.
17.	87.
18.	73.
19.	57.
20.	72.
21.	61.
22.	79.
23.	102.
24.	84.
25.	92.
26.	67.
27.	89.
28.	97.
29.	86.
30.	114.
31.	81.
32	94 
 }; 
      \addplot[red]  table[x=var1,y=var2] { 
      var1  var2 
0.	250.
1.	365.
2.	431.
3.	557.
4.	431.
5.	405.
6.	431.
7.	330.
8.	240.
9.	194.
10.	138.
11.	95.
12.	80.
13.	98.
14.	112.
15.	80.
16.	80.
17.	87.
18.	85.
19.	120.
20.	141.
21.	100.
22.	77.
23.	91.
24.	92.
25.	70.
26.	78.
27.	89.
28.	100.
29.	98.
30.	74.
31.	89.
32	105 
 }; 
      \addplot[blue]  table[x=var1,y=var2] { 
      var1  var2 
0.	250.
1.	256.
2.	171.
3.	149.
4.	110.
5.	88.
6.	111.
7.	101.
8.	105.
9.	71.
10.	97.
11.	96.
12.	96.
13.	90.
14.	69.
15.	73.
16.	140.
17.	93.
18.	96.
19.	69.
20.	81.
21.	77.
22.	83.
23.	90.
24.	101.
25.	77.
26.	82.
27.	96.
28.	80.
29.	71.
30.	85.
31.	107.
32	75 
 }; 
      \addplot[blue] table[x=var1,y=var2] { 
      var1  var2 
0.	250.
1.	156.
2.	161.
3.	109.
4.	89.
5.	99.
6.	107.
7.	100.
8.	77.
9.	65.
10.	61.
11.	81.
12.	82.
13.	72.
14.	80.
15.	58.
16.	97.
17.	66.
18.	70.
19.	71.
20.	77.
21.	73.
22.	106.
23.	112.
24.	97.
25.	79.
26.	101.
27.	109.
28.	114.
29.	110.
30.	94.
31.	87.
32	91
 }; 
      \addplot[red]  table[x=var1,y=var2] { 
      var1	var2
0.	250.
1.	279.
2.	355.
3.	424.
4.	497.
5.	356.
6.	388.
7.	314.
8.	276.
9.	179.
10.	129.
11.	94.
12.	104.
13.	108.
14.	91.
15.	76.
16.	98.
17.	69.
18.	74.
19.	72.
20.	62.
21.	89.
22.	99.
23.	116.
24.	114.
25.	69.
26.	83.
27.	84.
28.	101.
29.	103.
30.	95.
31.	70.
32	102
 }; 
      \addplot[blue]  table[x=var1,y=var2] { 
      var1	var2
0.	250.
1.	186.
2.	217.
3.	157.
4.	142.
5.	94.
6.	99.
7.	78.
8.	73.
9.	47.
10.	71.
11.	95.
12.	82.
13.	69.
14.	89.
15.	82.
16.	88.
17.	99.
18.	75.
19.	81.
20.	90.
21.	76.
22.	84.
23.	102.
24.	120.
25.	84.
26.	105.
27.	118.
28.	85.
29.	86.
30.	84.
31.	69.
32	72
 }; 
      \addplot[blue]  table[x=var1,y=var2] { 
      var1	var2
0.	250.
1.	202.
2.	142.
3.	90.
4.	98.
5.	121.
6.	102.
7.	89.
8.	95.
9.	68.
10.	82.
11.	71.
12.	92.
13.	103.
14.	89.
15.	107.
16.	75.
17.	97.
18.	110.
19.	85.
20.	88.
21.	91.
22.	88.
23.	104.
24.	85.
25.	76.
26.	85.
27.	85.
28.	98.
29.	74.
30.	98.
31.	73.
32	71
 }; 
      \addplot[blue]  table[x=var1,y=var2] { 
      var1	var2
0.	250.
1.	240.
2.	233.
3.	217.
4.	186.
5.	122.
6.	88.
7.	66.
8.	75.
9.	86.
10.	111.
11.	60.
12.	86.
13.	71.
14.	84.
15.	56.
16.	87.
17.	67.
18.	95.
19.	90.
20.	101.
21.	116.
22.	91.
23.	77.
24.	109.
25.	75.
26.	88.
27.	111.
28.	139.
29.	79.
30.	71.
31.	78.
32	84
 }; 
      \addplot[red]  table[x=var1,y=var2] { 
      var1	var2
0.	250.
1.	349.
2.	398.
3.	576.
4.	476.
5.	463.
6.	390.
7.	260.
8.	200.
9.	124.
10.	108.
11.	92.
12.	111.
13.	104.
14.	88.
15.	106.
16.	78.
17.	104.
18.	104.
19.	88.
20.	81.
21.	94.
22.	73.
23.	96.
24.	72.
25.	69.
26.	93.
27.	76.
28.	82.
29.	97.
30.	91.
31.	93.
32	109
 };
      \addplot[blue]  table[x=var1,y=var2] { 
      var1	var2
0.	250.
1.	237.
2.	204.
3.	175.
4.	93.
5.	79.
6.	109.
7.	96.
8.	80.
9.	69.
10.	94.
11.	68.
12.	74.
13.	64.
14.	89.
15.	51.
16.	86.
17.	77.
18.	83.
19.	67.
20.	59.
21.	83.
22.	88.
23.	90.
24.	93.
25.	109.
26.	91.
27.	86.
28.	104.
29.	95.
30.	59.
31.	103.
32	80
 };
    \end{axis}
  \end{tikzpicture}
 \caption{Trajectories of $X_1(t)$ for $n=32$ chains for $t\le 16$ for the 
   Schl\"ogl model in which only $X_1(t)$ varies (left) 
	 and for $n=16$ and $t\le 32$ when all copy numbers can vary (right).}
 \label{fig:schloegl-per-run}
\end{figure}

\red{The left panel of Figure~\ref{fig:schloegl-var-n} shows $\Var[\hat\mu_n]$ 
versus $n$ in log-log scale for the OSLAIF sort, for various point sets.  
The right panel shows $\Var[\hat\mu_n]$ as a function of $n$ under Sob+LMS, in a log-log-scale. 
 The estimated convergence rates $-\hat\beta$ are mostly between $-1.3$ and $-1.6$, 
which beats the MC rate of $-1$.
Notice the bump at $n = 2^{18}$ for the lattice rules.
It indicates that the selected lattice parameters for this $n$ are not ideal
for this specific example.  
When we made a search for parameters using order-dependent weights with $\rho=0.05$, 
as explained in Section~\ref{sec:rqmc-points}, the bump disappeared and 
the results were better. Since these weights give more importance to the 
low-dimensional projections, this suggests that the bump in the figure is due to 
a point set with one (or more) bad low-dimensional projection for this particular $n$.
In our reported results, we did not want to fine tune the parameters for each example and each $n$, 
because we think most users will not want to do that and it is not essential.}

One important observation is the large difference in MC variance between $T=4$ and $T=32$;
it is larger at $T=4$ by a factor of about 100.  
The mean $\EE[\hat\mu_n]$ also depends on $T$: it is about 240 at $T=4$ and about 86 at $T=32$.
\hflorian{FYI, the estimated mean at $T=4$ is 243.11 with $s=16$ steps and 239.4 with 
   $s=128$ steps. At $T=32$ it is 86.3.}%
What happens is that the trajectories have roughly two very different kinds of 
transient regimes between $t=0$ and about $t=10$.
For some trajectories, $X_1(t)$ goes up to somewhere between 400 and 600 at around $t=4$,
then goes down to around the long-term mean, say between 70 and 100.  
For other trajectories, $X_1(t)$ decreases right away to between 70 and 100 at around $t=5$.  
Figure~\ref{fig:schloegl-per-run} illustrates this behavior, with 16 sample paths.
\red{This behavior differs from that of the bistable system discussed before and in \cite{sBEE19a}.}
It explains the much larger variance at $T=4$ than at $T=32$ and it also shows 
why it is very hard to predict the state at some larger $T$ from the state at $t=1/4$,
say, hence the difficulty to estimate an ``optimal'' importance function.
Despite this, Array-RQMC performs quite well with simple sorts and brings large 
efficiency improvements compared with MC and RQMC.
\hpierre{Time $t=1/4$ is the first step.  I understand from the picture that the state at any
 time $t < 10$, say, gives little information on $X_1(32)$.  
 This means that it would be very hard to estimate the parameters of a SDIF for $T=32$.   
 It might explain why the OSLAIF does not perform as well for $T=32$ as for $T=4$. 
 On the other hand, the values
 of $X_1(t)$ for $t=1/4$ or $t=1$ for example, could already indicate whether we have a 
 blue or red trajectory.  So for $T=4$, it seems that there is hope.
 But that does not mean that the sort must be step-dependent.}%

\begin{table}[!htbp ] 
	\centering
	\footnotesize
	\caption{Estimated rates $\hat\beta$, $\vrf$, and $\eif$ for the Schl{\"o}gl system,
	         with four types of sorts for Array-RQMC.}
\label{tab:schloegl} 
 \begin{tabular}{ c|l| crr | crr | crr }
\hline
          & & \multicolumn{3}{c|}{$T=4,\ s=16$} & \multicolumn{3}{c|}{$T=4,\ s=128$} 
						& \multicolumn{3}{c}{$T=32,\ s=128$} \\
\hline						
 $\EE[g(\bX_s)]$ & &\multicolumn{3}{c|}{243}   &\multicolumn{3}{c|}{239} 
	           &\multicolumn{3}{c}{86}  \\
\hline						
 MC $\Var$ & &\multicolumn{3}{c|}{27,409}   &\multicolumn{3}{c|}{27,471} 
	           &\multicolumn{3}{c}{270}  \\
\hline
 Sort & Sample  & $ \hat\beta $ & $\vrf$ & $\eif$  & $ \hat\beta $ & $\vrf$ & $\eif$
                & $ \hat\beta $ & $\vrf$ & $\eif$\\
\hline
      & MC			& 1.00	& 1		& 1		& 1.00	& 1	& 1	& 1.00	& 1		& 1    \\
      & RQMC	  & 1.14	& 11	& 12	& 1.04	& 7	& 8	& 1.29	& 211	& 203 \\
\hline 
\multirow{4}{*}{by $S_1$}	
		&	Lat+s 			& 1.54	& 2283	& 2099	& 1.04 	& 2003	& 1406	&  1.01	& 255	& 221 \\
		&	Lat+s+b 		& 1.24	& 4385	& 4028 & 1.10	& 1596	& 1121	& 1.02 	& 189	& 159 \\
		&	Sob+LMS 		& 1.41	& 5835	& 5500	& 1.38	& 1760	& 1332	& 1.05 	& 268	& 191 \\
%
\hline			
\multirow{4}{*}{OSLAIF}	
		&	Lat+s 			& \bf{1.58}	& 2686	& 1477	& 1.12 	& \bf{3637}	& \bf{2201}	& 1.08 	& 366	& 406 \\
		&	Lat+s+b 		& 1.24	& 4385	& 3901 & 1.08	& 1464	& 974	& 1.08 	& 442	& 403 \\
		&	Sob+LMS 		& 1.35	& 5823	& 5329	& \bf{1.47}	& 3215	& 2187	& 1.10 	& 666	& 525 \\
\hline
%
%
%
		\multirow{4}{*}{Batch}
		&	Lat+s	 		& 1.55	& 1283	& 1144	& 1.42	& 906	& 342	& 1.20 	& 539	& 573 \\
		&	Lat+s+b 		& 1.38	& 4077	& 3633	& 1.23	& 930	& 522	& \bf{1.29} 	& 1440	& \bf{1582} \\
		&	Sob+LMS 		& 1.46	& \bf{6434}	& \bf{5760}	& 1.41	& 1847	& 1105	& 1.27 	& 1569	& 1200 \\
		\hline
		\multirow{4}{*}{Hilbert}
		&	Lat+s	 		& 1.35	& 990	& 818	& 1.17	& 508 	& 274	& 1.04 	& 1151	& 850 \\
		&	Lat+s+b			& 1.28	& 3157	& 2610 	& 0.88	& 337 	& 179	& 0.93 	& 600 	& 438 \\
		&	Sob+LMS 		& 1.55	& 3512	& 3138	& 1.23	& 534	& 321	& 1.28	& \bf{1611}	& 1221 \\
		\hline
	\end{tabular} 	
\end{table}

\subsection{A model of cyclic adenosine monophosphate activation of protein kinase A}

This example is a model for the cyclic adenosine monophosphate  ($\camp$) activation of
protein kinase A  ($\pka$), taken from \cite{sKOH12a} and \cite{sSTR15a}. 
This model is interesting because it has $\ell=6$ and $d=6$, which are both larger than 
in the previous examples. 
The six molecular species $S_1$ to $S_6$ are (in this order) $\pka$, $\camp$, 
the partially saturated $\pkatwo$, the saturated $\pkafour$, 
the \red{regulatory} subunit $\pkar$, and the \red{catalytic} subunit $\pkac$. 
\hflorian{Interchanged the order of $\pka$ and $\camp$ as well as of $\pkar$ and $\pkac$.
This corresponds to the order in which we list them in the reaction scheme.}%
The $d=6$ possible reactions are depicted here:
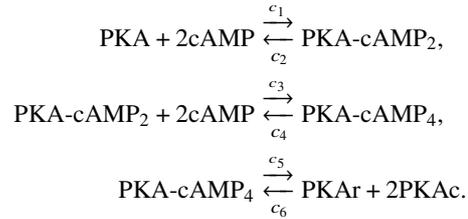
\begin{align*} 
	\pka+2\camp & \stackrel{\xrightarrow{c_1}}{\xleftarrow[c_2]{}}  \pkatwo ,  \\
	 \pkatwo +2\camp  &\stackrel{\xrightarrow{c_3}}{\xleftarrow[c_4]{}} \pkafour,  \\
	 \pkafour  & \stackrel{\xrightarrow{c_5}}{\xleftarrow[c_6]{}} \pkar+2\pkac.
\end{align*} 
The reaction rates are 
$ c_{1} = 2.6255\times10^{-6}$, $ c_{2} = 0.02$, $ c_{3} = 3.8481\times10^{-6}  $, 
$ c_{4} = 0.02 $,  $ c_{5} = 0.016 $ and $c_{6} = 5.1325\times10^{-5} $. 
We simulate this system with the same parameters as \cite{sPAD16a}, except that we 
assume that the molecules are homogeneously distributed in the volume 
\red{(so the example fits our framework)} and we choose 
a fixed $\tau$ as opposed to changing it adaptively.  
%
\hflorian{Converted rates corresponding to \cite{sPAD16a}.}%
\hflorian{Should we include units or not? 
  Do the rates \emph{with} units mean the same as those \emph{without} units?}%
At time zero there are 33,000 molecules of $\pka$, 33,030 molecules of $\camp$,
and 1,100 molecules of each other species.  
We take $T=0.05$ and $\tau = T/256$, so we have $s= 256$ steps. 
This problem requires RQMC points \new{in 7 to 12} dimensions with Array-RQMC,
\new{depending on the sort},
compared to 1536 dimensions with classical RQMC.

We report experiments with two different objective functions.
The first one is $\EE[X_1(T)]$, the expected number of molecules of {\pka} at time $T$,
and the second one is $\EE[X_5(T)]$, the expected number of molecules of {\pkar} at time $T$.
In each case, we implemented and tested the OSLAIF and SDIF methods to select a mapping $h$ 
to one dimension.  We also tried the multivariate batch sort and
the Hilbert sort from Section~\ref{sec:sorting-strategies}.
The best performers were the OSLAIF map, the batch sort, and the Hilbert sort.
\hflorian{For the batch sort, the batch sizes were selected using the \emph{proportion exponents} rule
explained in the class \texttt{BatchSort} of SSJ \cite{iLEC16j}, with exponent $1/6$.
Specifically, whenever there are $n$ elements to sort, they are partitioned into 
$\lceil n^{1/6}\rceil$ batches (or the nearest integer).  For example, if we start with 
$n=2^{19}$, the $n$ points will first be split into $n_0 = \lceil 2^{19/6}\rceil = 10$ batches
at the first level, then each batch is split into $n_1 = \lceil (2^{19}/10)^{1/6}\rceil = 7$ batches 
at the second level, then 5, 4, 3, and 3 batches at the higher levels.}%
\hflorian{Moved the discussion about batch sort further down}
\hflorian{I took the version of the batch sort which uses proportional exponents 
  $\boldsymbol\alpha=\{0.17,0.17,0.17,0.16,0.16\}$.}
\hpierre{In the batch sort, the ordering of the coordinates is very important.  One should 
  sort first by the most important coordinate, then the second most important one, etc.
  Here it seems that you have not done that, but just took $x_1$, then $x_2$, etc.
	In this example, for the second objective function, 
	it seems $x_5$ is the most important, so I would first sort on $x_5$,
	perhaps in at least $n^{1/2}$ batches, then perhaps sort the batches on $x_6$,
	which might be the second most important coordinate, etc.  I think you should try a larger
	number of batches for the first coordinates.  Also, you do not have to use all coordinates.
	You can use only the first 2 or 3 coordinates for the sort.  
	Using only $x_5$ for the batch sort with $n_0 = n$ batches at the first level 
  gives the special case that you have as the first entry in the table.
	For the first objective function, I don't know which are the most important variables,
	but when looking at the OSLAIF, it might be $x_1$, then $x_2$ (or vice-versa), then $x_3$.
	I think it is important to check that, to have an effective batch sort.}

\red{For $g(\bx)=x_1$ the OSLAIF} is given by the polynomial $h(\bx)=x_1+\tau(-c_1 x_1 x_2(x_2-1)/2 + c_2 x_3)$.
In this function, $x_1$ outweighs the term $-\tau c_1x_1x_2(x_2-1)/2$
on average, followed by $\tau c_2 x_3$.  
This suggests taking $x_1$ as the most important 
coordinate for the sort, followed by $x_2$ and $x_3$.
For the batch sort, we used these three coordinates in this order,
with batch sizes $n_j=\lceil n^{\alpha_j}\rceil$. 
We first tried $(\alpha_1,\alpha_2,\alpha_3) = (1/2, 3/8, 1/8)$, 
but $(\alpha_1,\alpha_2) = (1/2, 1/2)$ performed slightly better and is used for our results.
\hpierre{We tried other values and some gave better results for some point sets, e.g., $(1/4, 1/2, 1/4)$ 
gave a \vrf{} of 3656 for Lat+s.} 
\hpierre{We may also just leave out this sentence.  
 This is probably getting too much into the details, and it does not help much the users.}
\hflorian{That's fine with me. Perhaps we can keep it in the response to the referee. I can't 
tell from experience how elaborate they want the answers to be.}
\hflorian{The referee also asks us about applying the random forest technique to this example.
We might not put it in the paper (it doesn't add any value), but we must address it in the 
response letter. See my notes, first item on p. 27. Moreover, he asks about the cost and the
number of samples required for the random forest techniuqe (also not really paper material), 
which is discussed in the next item in my notes.}

Table~\ref{tab:pka} summarizes our results for $g(\bx) = x_1$ (the {\pka} case).
The estimated mean and variance per run with MC are 19663 and 1775, respectively.
The batch sort with Sob+LMS gives the largest improvement empirically.
Classical RQMC also performs surprisingly well despite the large number of dimensions.
\hpierre{Referee 2 asks why... }
With Array-RQMC, we also observe empirical convergence rates $\hat\beta$ consistently
better than the MC rate of 1.0. This indicates that the VRF should increase further with $n$.

Table~\ref{tab:pkar} gives the results for $g(\bx) = x_5$ (the {\pkar} case).
The estimated mean and variance per run with MC are about 716 and 47, respectively.
The OSLAIF is given by $h(\bx) = x_5 + \tau (c_5 x_4 - 0.5 c_6 x_5 x_6(x_6-1))$.
Given that $x_4$, $x_5$, and $x_6$ remain roughly between 500 and 1000 in this model,
and that $\tau = 1/5120$, the dominating term in this function is (by far) $x_5$, followed by 
$ -\tau c_6 x_5 x_6^2  \approx - 2.5\times 10^{-3} x_5$. 
Based on this, for the batch sort, we initially used the coordinates $x_5, x_6, x_4$ in this order.  
\red{For the reported results, we took $(\alpha_5,\alpha_6,\alpha_4) = (1/2, 1/4, 1/4)$
for the batch exponents.  We tried other choices such as  
$(\alpha_5,\alpha_6,\alpha_4) = (1/2, 3/8, 1/8)$, $(1/3, 1/3, 1/3)$, etc.,
and similar results were obtained, but with weaker figures for Lat+s.}

We also tried SDIF with various types of functions, but it did not really perform better.
While doing this, we applied \red{a procedure based on} the random forest permutation-based 
statistical method of \cite{tBRE01a} to detect the most important variables in a noisy function.
This procedure told us that $x_6$ was the most important variable for the sort, at all steps. 
Based on this, we also tried sorting the states by $x_6$ (the number of $\pkac$ molecules) only. 
This is a degenerate form of batch sort with $n_6=n$.
We call it ``By $\pkac$'' in Table~\ref{tab:pkar}.
\hflorian{I had explained that in previous versions of the manuscript. I took it from 
\url{https://doi.org/10.1111/gcb.13666}, Section 2.5, end of 3rd page. They don't 
mention any specific name of the method either\ldots In all brevity: I fitted the 
states of $n=2^{19}$ states at some step to the final $g(\bx)$ via a generic multivariate
polynomial of degree 3 and stored the coefficients in $\bh$. Then I permuted the states, but
kept the responses in their original order. I fitted again a polynomial of degree 3 to the 
permuted states and stored the results in $\tilde\bh$. Afterwards, I computed the correlation
between $\bh$ and $\tilde\bh$ component wise. If the correlation was close to one, I considered
the corresponding term unimportant, if it was small, I considered the term important. I did that
for several different steps and permutations and the answer was always the same.}%
\hflorian{We performed $n=2^{19}$ independent runs to estimate a good $h_j$ as described in 
Section~\ref{sec:sorting-strategies}.  
For each step $j$, we fitted a generic multivariate polynomial $h_j$ in six dimensions,
of the form 
$$
  h_j(x_1,x_2,\dots,x_6) 
	 = \sum_{0\leq\varepsilon_j\leq 3} \beta_{\varepsilon_1,\dots,\varepsilon_6} 
	     x_1^{\varepsilon_1} x_2^{\varepsilon_2} \cdots x_6^{\varepsilon_6},
$$
to the copy numbers of $\pkar$ of the chains at time $T$. 
The values of $h_j$ evaluated at the states of each chain were stored in a vector $\bh$. 
Subsequently, we applied the following sensitivity analysis procedure to assess the importance
of each of the six variables.
We permuted the data values of one specific projection $x_1^{\varepsilon_1}x_2^{\varepsilon_2}\cdots x_6^{\varepsilon_6}$ by some permutation $\pi$ and fitted a polynomial $\tilde{h_j}$ to this falsified data.  The values of $\tilde{h_j}$ evaluated at the states of each chain again were stored in a vector $\tilde\bh$. As a measure of importance we considered one minus the correlation of $\bh$ and $\tilde\bh$, i.e., the closer the correlation of $\bh$ and $\tilde\bh$ was to zero, the more important we considered the specific projection. We applied this procedure to every projection with different (randomly picked) permutations $\pi$ and, as it turned out, the cubic term in the copy number of $\pkac$ was the only important projection for this simulation with an importance of over 0.99 in the above sense for every $\pi$ we have tried.}

The OSLAIF, Batch, and ``By $\pkac$''  sorts perform similarly.
They outperform the Hilbert sort and also classical RQMC.
Their empirical convergence rates $\hat\beta$ are also significantly larger than 1.
This example illustrates two facts. 
First, the dimension of the state is not the ultimate criterion for Array-RQMC to perform well.
Secondly, customizing sorting algorithms based on information on the underlying model can 
improve results significantly.

Following a request from one referee, we performed experiments in which we estimated
the second and third moments of the $\pka$ and $\pkac$ copy numbers at time $T$, using both
OSLAIF and a batch sort. The OSLAIF is easily computed because the second and third moments
of the Poisson distribution are known explicitly.
With OSLAIF, the \vrf{} with Array-RQMC was around 900 to 1200 with 
the various point sets.  The batch sort with Lat+s gave the best performance, 
with a \vrf{} around 3600.
We were also asked to run experiments in which we estimate several expectations simultaneously
using the same runs (and therefore the same sort).   Our software does not allow this but we 
``simulated'' it by running it for different expectations with the same sort.
\new{We estimated the mean as well as the second and third moments of $\pka$ copy numbers with
the OSLAIF for $g(\bx)=x_1$ and obtained the same $\vrf$ as in Table~\ref{tab:pka}.}
When we estimated the expected copy numbers of all six types of molecules using the same sort,
for this example, the average \vrf{} was reduced by a factor of about 2 compared with the case
where we have a sort adapted to each expectation. So this is still reasonably effective. 
\new{For some coordinates, the \vrf{} remained the same while it dropped by factors between 10 to
30 for the worst ones. This can certainly be improved by selecting the sort more carefully,
but we intentionally restrained ourselves to simple methods that are easy to apply.}
The sorts we used were an ``average OSLAIF'', which takes the average of the six OSLAIF 
functions $h$ adapted to each of the six expectations, and a batch sort with batch exponents
all equal to 1/6 \new{with the state variables kept in their default order.}
\hpierre{For both $\pka$ and $\pkac$?}%

\hpierre{This means the OSLAIF misses completely the most important variable, which is $x_6$,
  and still it performs almost as well.  Not obvious to explain, 
	but it suggests that there can be several effective ways to sort in general.}%
\hpierre{Given that $x_6$ is the most important variable, the most natural thing to do 
  would be to put it as the first variable in the batch sort.  
	That is, sort based on $x_6, x_5, x_4$, in this order, with $n_1 = n^{1/2}$
	or  perhaps even $n_1 = n^{2/3}$. Makes sense?  
	If it works well, we can report the batch sort with the previous ordering and also this one.}%
\hflorian{None of the settings we tried for the batch sort, which were adapted to the OSLAIF,
seems to capture the additional information that we captured by the permutation-based statistical
procedure which we have used for the ``By $\pkac$'' sort.}
\hflorian{I rephrased it, to avoid ``SDIF data''. FYI: I meant the data of $2^{19}$ chains simulated
by MC, which we also used to construct the SDIF.}
%
\hpierre{The ``By $\pkac$'' is just a special case of the batch sort.  
 The batch sort is mush more flexible, so it should perform at least as well, 
 and therefore much better than what we see, if done right. 
 So there seems to be a problem with the ordering of variables, or something else.}
\hflorian{Yes. But there are several heuristics how we could select batch sizes. We 
 are stating that OSLAIF-based choices are not that great,
 while the additional knowledge we use for the ``By $\pkac$'' gives a good choice.}
\hflorian{As indicated way further above: The referee asks for 2 more things here: 
1) using non-linear $g$ and 2) simultaneous estimation of \emph{all} copy numbers. 
We mention that we will do 1) already in Section 2 with 2nd and 3rd moments, but 
haven't included anything. Details are in my notes, Section 1.2.2. For 2) he asks
if we can find a sort that works for all copy numbers at once. The sum of OSLAIFs
worked comparably well (see my notes Section 1.2.3).}

\begin{table}[!htbp ] 
	\centering
	\footnotesize
	\caption{Estimated rates $\hat\beta$, $\vrf$, and $\eif$, for $\pka$ with $T=0.05$, $s=256$.}
	\label{tab:pka} 
	\begin{tabular}{|c|l|c|r|r|}
		\hline
		Sort & Sample 			& $\hat\beta$	& \multicolumn{1}{c|}{\vrf}	& \multicolumn{1}{c|}{\eif}  \\
		\hline
		&	MC  			& 1.00		& 1 	& 1\\
		& RQMC			& 1.08		& 464	& 603 \\
 		\hline
		\multirow{4}{*}{OSLAIF}	
		&	Lat+s		 		& \bf{1.44}		& 1141	& 671 \\
		&	Lat+s+b		 		& 1.25		& 830 	& 460 \\
		&	Sob+LMS 			& 1.28		& 1112	& 762 \\
		\hline
%
%
%
%
%
		\multirow{4}{*}{Batch} 
		&	Lat+s		 		& 1.30		& 1535	& 801	\\
		&	Lat+s+b		 		& 1.09		& 1446 	& 681 \\
		&	Sob+LMS 			& 1.17		& \bf{1979} 	& \bf{1146} \\
		\hline
		\multirow{4}{*}{Hilbert}
		&	Lat+s		 		& 1.25		& 1054	& 400 \\
		&	Lat+s+b		 		& 1.17		& 855	& 305 \\
		&	Sob+LMS 			& 1.19		& 1258	& 545 \\
		\hline
%
%
\end{tabular} 	
\end{table}

\begin{table}[!htbp ] 
	\centering
	\footnotesize
	\caption{Estimated rates $\hat\beta$, $\vrf$, and $\eif$, for $\pkar$ with $T=0.05$, $s=256$.}
	\label{tab:pkar} 
	\begin{tabular}{|c|l|c|r|r|}
		\hline
		Sort & Sample 			& $\hat\beta$	& \multicolumn{1}{c|}{\vrf}	& \multicolumn{1}{c|}{\eif}  \\
		\hline
		&	MC 					& 1.03		& 1 	& 1\\
		&	RQMC			 & 1.17		& 39	& 45 \\
 		\hline
		\multirow{4}{*}{OSLAIF}	
		&	Lat+s 				& 1.42		& 3634	& 1727\\
		&	Lat+s+b 			& 1.38		& 1491 	& 673  \\
		&	Sob+LMS 			& 1.47		& 2062	& 1163 \\
		\hline
%
%
%
%
		\multirow{4}{*}{Batch}  
		&	Lat+s		 		  & 1.54		& \bf{3961}	& \bf{2104} \\
		&	Lat+s+b		 		& 1.44		& 1416 	& 728 \\
		&	Sob+LMS 			& \bf{1.65}		& 1224 	& 811 \\
		\hline
%
%
		\multirow{4}{*}{By $\pkac$}	
		&	Lat+s 				& 1.33		& 2470	& 1513 \\
		&	Lat+s+b 			& 1.36		& 1364 	& 779 \\
		&	Sob+LMS 			& 1.45		& 1856 	& 1386 \\
		\hline
%
		%
		\multirow{4}{*}{Hilbert}
		&	Lat+s		 		& 1.17		& 135	& 54 \\
		&	Lat+s+b		 		& 1.12		& 88 	& 27 \\
		&	Sob+LMS 			& 1.24		& 126 	& 60 \\
		\hline 
	\end{tabular} 	
\end{table}

\subsection{Quasi-steady state approximation examples}

Quasi-steady state approximation (QSSA) is a simplification approach to reduce the size
of a model so that it can be simulated much faster \citep{sCAO05b,sKIM15a,sRAO03a,sTHO12a}.
It applies in situations where some of the reaction types occur at a slow time scale, 
whereas other reaction types occur at a much faster time scale.  
In the simplified model, one assumes that after each slow-type reaction, 
a very large number of the fast-type reactions occur in an infinitesimal time period,
so that the system reaches steady-state very quickly with respect to those reactions.
One then assumes that until the next slow-type reaction, the state of the vector of variables 
that are affected only by the fast-type reactions is distributed according to its steady-state 
conditional distribution given the other variables (which we call the slow-type variables
and are assumed fixed).
Under these assumptions, only the slow reactions need to be simulated, using propensities 
that are functions of the states of the slow-type variables only.
These functions are often non-polynomial.
The validity of this type of approximation is studied in many papers, 
including \cite{sKIM15a}, \cite{sTHO12a}, and other references cited there.
Here, we focus on assessing how Array-RQMC can improve the statistical efficiency 
for simplified models in which the reaction rate functions depart from the mass action kinetics.
To fit our framework, we combined QSSA with tau-leaping.
\hpierre{The approximation should remain approximately valid if we take a very small $\tau$, 
so there is rarely more than one reaction per step.}

One of the reviewers suggested the following model of \emph{cooperative enzyme kinetics}, 
given in Eq.~(9) of \cite{sKIM15a}.  
\hflorian{Actually, the referee asks us to look at the OSLAIF when we deviate from mass action
kinetics. As only the propensity functions change, the OSLAIF can be constructed as before. We
only have to keep in mind that the mean of the Poisson variates ($\tau$ times propensity evaluated
at the current state) looks different in this case. }
Its simplified version has reaction rates 
given in their Eq.~(10),
with two state variables $S$ and $P$ which correspond to our $S_1$ and $S_2$.
\new{The reactions can be depicted as $\emptyset \xrightarrow{c_1}  S_{1}$ and  
$S_{1} \xrightarrow{c_1} S_{2}$, with propensities $a_1(\bx)=1$ and 
$a_2(\bx)= x_1^2/ (K_m^2 + x_1^2)$, respectively, and constants $c_1=0.5,c_2=1$, and  
$K_m = 2.02\times 10^5$. }
%
We also took $T=2^{17}$ with $s=1024$, so $\tau=2^7 = 128$, and we started with an empty system.

We consider two cases: (1) when we want to estimate $\EE[X_1(T)]$ and
(2) when we want to estimate $\EE[X_2(T)]$.
In case (1), the expectation does not depend on the current number of molecules of $S_2$, 
so we have a one-dimensional chain only, and the sorting is easy.
In case (2) the expectation depends on both numbers of molecules, so the state is two-dimensional.
For this second case, the OSLAIF gives 
$h(\bx)=x_2 + \tau a_2(\bx) = x_2 + \tau k_p x_1^2 / (K_m^2+x_1^2)$
and we use batch exponents $\balpha=(1/2,1/2)$ for the batch sort.
The results are reported in Table~\ref{tab:enzyme2D}.
We see that Array-RQMC can provide very large gains, much larger than classical RQMC.
For case (2), we find that $x_1$ is the most important variable for the sort:
sorting by the copy number of $S_1$, or a batch sort that takes $x_1$ as the first variable,
give the best results.  The OSLAIF is not competitive in this case because as time goes on,
$x_2$ increases, and $h(\bx)$ does not give enough weight to $x_1$ compared with $x_2$.
In this system, it takes a very large pool of $S_1$ to start producing $S_2$ at a 
significant rate, and this is why $x_1$ is important.
\hpierre{Perhaps a good predictor of $\EE[X_2(T)]$ could be the current sum $X_1(t) + X_2(t)$.}
\hflorian{That makes sense and I will try it. Hopefully time allows us to finish it in time. 
In the paper we could ultimately spend 1-2 lines on that, if it works out.}

\begin{table}[!htbp] 
\centering
\caption{Estimated rates $\hat\beta$, $\vrf$, and $\eif$, for the enzyme kinetics example, 
  with $T=2^{17}$ and $s=2^{10}$, for $g(\bx)=x_1$ (left) and for $g(\bx)=x_2$ with Lat+s (right).}
\label{tab:enzyme2D}
\small
\begin{center}
\hbox{\hfill
\begin{tabular}{l|  crr}
\hline
 $\EE[X_1(T)] $  &\multicolumn{3}{c}{61,512} \\
 \hline
 MC $\Var$ &\multicolumn{3}{c}{55,398}     \\
\hline
Point sets & $ \hat\beta $ & $\vrf$ & $\eif$ \\
\hline
MC		    & 1.00	& 1      & 1   	 \\
RQMC        & 1.05  & 4,532       & 4,857     \\
Lat+s	    & \bf{1.92}	& 57,267   & 30,499  \\
Lat+s+b		& 1.51	& 75,809   & 42,319   \\
Sob+LMS		& 1.55	& \bf{129,414} & \bf{79,531} \\
\hline
\end{tabular} 	
\hspace{30pt}
\begin{tabular}{l|  crr}
\hline		
 $\EE[X_2(T)] $  &\multicolumn{3}{c}{4,024} \\
 \hline
 MC $\Var$ &\multicolumn{3}{c}{4,479}     \\
\hline
Point sets & $ \hat\beta $ & $\vrf$ & $\eif$ \\
\hline
MC		    & 1.00	& 1      & 1   	 \\
RQMC        & 1.04  & 365     & 387     \\
OSLAIF	    & 1.43	& 1,469   & 708  \\
Batch		& \bf{1.81}	& 14,570   & 7,250   \\
by $S_1$	& 1.70	& \bf{20,153} & \bf{12,344} \\
by $S_2$	& 1.00	& 273 &  165 \\
\hline
\end{tabular}}	
\end{center}
\end{table}

\hpierre{I am not convinced that we should add the ``Coupled Flows'' example here.
 I do not see how it would make our message stronger, 
 especially that we do not compare with the approach 
 of Hellander and also that the \eif{} are not much better than for RQMC.
 We are also short on space. I would leave it out for this paper. }
\hflorian{I understand your point. My concern was only the referee. As mentioned above
he actually wanted to see something on the OSLAIF for non-polynomial propensity functions.
It depends how we interpret is question. If he wants to see how the OSLAIF can be constructed, 
we might just add 1-2 sentences (as indicated above), and we're good. If he wants to see how 
the OSLAIF performs, we might need to add something explaining that there are certainly 
situations in which the combination of the OSLAIF with, e.g., rational propensities performs well,
but this just isn't one of those cases.}

\color{black}
\section{Conclusion}

We have studied the combination of the fixed step $\tau$-leap algorithm with Array-RQMC for well-mixed chemical reaction networks and found that in this way, we can reduce the variance in comparison to MC significantly.  In contrast to the simulation with traditional RQMC, this approach could often also improve the convergence rate of the variance. Array-RQMC requires to sort the chains by their states at each step of the chain. This can be done with a multivariate sort.  But we also showed that one can construct sorts by mapping the states into the real numbers via a simple importance function, and then the sorting is trivial. 
\hflorian{Despite the fact that standard multivariate sorts can become costly when the state space has large dimensionality, we observed that their combination with array-RQMC, and usually even classical RQMC, is more efficient than MC.}%
Some basic knowledge of how the model behaves is of course useful to identify the 
important state variables that should be retained for a batch sort or to build a better 
importance function, which in turn can improve the convergence rate of the variance.
\new{In our experiments, Array-RQMC was never worse than MC, for all sorting methods.}
In follow-up work, it would be interesting to explore how automatic learning methods 
could be used to find better importance functions.
\hpierre{I think it is not very clear from our examples how the important variables
 can be identified easily, other than by trial and error, and some heuristics that do not
 always perform well.  On the other hand, we always manage to get important VRFs.
 So I think I need to reformulate this part.}

\begin{acknowledgements}
This work has been supported by a Canada Research Chair, an IVADO Research Grant, and an NSERC Discovery Grant number RGPIN-110050 to P. L'Ecuyer. F. Puchhammer was also supported by Spanish and Basque governments fundings through BCAM (ERDF, ESF, SEV-2017-0718, PID2019-108111RB-I00, PID2019-104927GB-C22, BERC 2018e2021, EXP. 2019/00432, ELKARTEK KK-2020/00049), and the computing infrastructure of i2BASQUE academic network and IZO-SGI SGIker (UPV).
\end{acknowledgements}

\bibliographystyle{spbasic}      
\bibliography{prob,math,fin,ift,optim,random,simul,stat,vrt}


\end{document}